\documentclass[a4paper,10pt,twoside]{cpc-hepnp}

\usepackage{multicol}
\usepackage{graphicx}
\usepackage{booktabs}
\usepackage{amssymb,bm,mathrsfs,bbm,amscd}
\usepackage[tbtags]{amsmath}
\usepackage{lastpage}
\usepackage{array,multirow,graphicx}

\begin{document}

\fancyhead[c]{\small Chinese Physics C~~~Vol. xx, No. x (201x) xxxxxx}
\fancyfoot[C]{\small 010201-\thepage}

\footnotetext[0]{}

\title{Spectroscopy and Decay properties of the charmonium}

\author{%
    Virendrasinh Kher $^{1}$,$^{2}$\email{vhkher@gmail.com}%
\quad Ajay Kumar Rai$^{2}$*\email{raiajayk@gmail.com}%
}
\maketitle

\address{%
$^1$ Applied Physics Department, Polytechnic, The M.S. University of Baroda, Vadodara 390002, Gujarat, India \\
$^2$ Department of Applied Physics, Sardar Vallabhbhai National Institute of Technology, Surat 395007, Gujarat, India \\
}

\begin{abstract}
The mass spectra of charmonium are investigated using a Coulomb plus linear (Cornell) potential. Gaussian wave function in position space as well as in momentum space are employed to calculate the expectation {\bf value} of potential and kinetic energy respectively. Various experimental states ($X(4660)(5^{3}S_{1})$, $X(3872)(2^{3}P_{1})$, $X(3900)(2^{1}P_{1})$, $X(3915)(2^{3}P_{0})$ and $X(4274)(3^{3}P_{1})$ etc.) are assigned as charmonium states. We also study the Regge trajectories, pseudoscalar {\bf and} vector decay constants, the Electric {\bf and} Magnetic dipole transition rates and the annihilation decay width for charmonium states.
\end{abstract}



\footnotetext[0]{\hspace*{-3mm}\raisebox{0.3ex}{$\scriptstyle\copyright$}2013
Chinese Physical Society and the Institute of High Energy Physics of the Chinese Academy of Sciences and the Institute of Modern Physics of the Chinese Academy of Sciences and IOP Publishing Ltd}%

\begin{multicols}{2}

\section{Introduction}
\label{intro}

 The discovery of the $J/\psi$, first bound state of $c$ and $\overline{c}$ quarks, known as charmonium, is published in Ref.\cite{Aubert:1974}, whereas Ref.\cite{Augustin:1974} describes the first observation of the  $\psi(2S)$  and marked the field of hadron spectroscopy with the beginning of an important testing ground for the properties of the strong interaction using QCD. Charmonium system allows the prediction of some of the parameters of the states, using non-relativistic
and relativistic potential models, lattice QCD, NRQCD and sum rules \cite{Brambilla2011}. Although the first charmonium state was discovered in 1974, there are still many puzzles in charmonium physics. The charmonium spectroscopy below the open charm threshold has been well measured and agrees with the theoretical expectations, however, there {\bf are} still lack of adequate experimental {\bf informations} and solid theoretical inductions for the charmonium states above the open charm threshold \cite{PDGlatest}. Recently many other new  resonances named $X\,Y\,Z$ particles have been discovered and are still under examination as these states do not match the predictions of the non-relativistic or semi-relativistic $q\bar{q}$ potential models.

In 1976, Siegrist and others at MARK-I collaboration (SLAC) observed the resonance $\psi(4415)$ with mass $4415\pm7$ MeV \cite{Siegrist:1976}. In 1978, DASP collaboration observed {\bf peaks} for $\psi(4040)$, $\psi(4160)$ and $\psi(4415)$ resonances with mass $4040\pm10$, $4159\pm20$ and $4417\pm10$ MeV respectively using non-magnetic detector \cite{Brandelik:1978}. Ablikim and others at BES collaboration and Mo and others at Beijing Institute HEP, determined the resonance parameters for $\psi(4040)$, $\psi(4160)$ and $\psi(4415)$ charmonium. Eichten identified that {\bf these} three resonances are $3^3S_1$, $2^3D_1$ and $4^3S_1$ with linear plus Coulomb potential model \cite{Eichten1980} and most later potential model calculations {\bf are agreed} with their identification. Recently, the LHCb collaboration  measured the mass  $4191_{-8}^{+9}$ MeV of the resonance $\psi(4160)$ with  $J^{PC} = 1^{--}$ \cite{Aaij2013}. In the year 2007, a resonant structure was observed by Belle collaboration with mass $4664\pm11\pm5$ MeV \cite{Wang:2007} and after one year later same collaboration observed a clear peak in the $e^+e^-\rightarrow \Lambda_c^+\Lambda_c^-$ invariant mass distribution and assumed that the observed peak to be a resonance {\bf of} mass $4634_{-7}^{+8}{}_{-8}^{+5}$ MeV with  the possibility of $5^3S_1$ charmonium state \cite{Pakhlova:2008}.
 
Rapidis and others at SLAC, LGW collaboration,  observed a resonance with mass $3772\pm6$ MeV, just above the threshold for the production of charmed particles \cite{Rapidis:1977}. In parallel observation, W. Bacino and others at SLAC  discovered and confirmed the  $\psi (3770)$ resonance with mass $3770\pm6$ MeV \cite{Bacino:1977} and the parameters were determined by SLAC and LBL collaborations \cite{Abrams:1979}. In 2006 BES Collaboration measured the precise measurements of the mass of $\psi (3770)$ resonance \cite{Ablikim:2006} and recently its   parameters have been measured using the data collected with the KEDR detector \cite{Anashin:2011}. The Belle collaboration reported the first observation of a new charmonium-like state with mass $3943\pm6\pm6$ MeV  in the spectrum of masses recoiling from the $J/\psi$ in the inclusive process $e^+e^-\rightarrow J/\psi + \text{anything}$,  and denoted it as $X(3940)$ \cite{Abe:2007a}. Later on, new measurement for the $X(3940)$ was performed  by the same collaboration and the mass $3942_{-6}^{+7}\pm6$ MeV was reported \cite{Abe:2007b}. The $3^1S_0$ state can be a
good candidate of the $X(3940)$ resonance \cite{Sreethawong:2013,Wang:2016}.

Evidence of a new narrow resonance $X(3823)$ was  found by Belle \cite{Bhardwaj:2013}, with its mass near to potential model expectations for the centroid of the $1^3D_J$ states. Recently, BESIII Collaboration \cite{Ablikim:2015a}, observed a narrow resonance $X(3823)$ through the process $e^{+}e^{-}\rightarrow \pi^{+}\pi^{-}X(3823)$ and confirmed that it is a good candidate for the $\psi(1^{3}D_{2})$ charmonium state. 

In year 2003, Belle Collaboration observed charmonium like state in the decay process $B^{\pm} \rightarrow K^{\pm}\pi^+\pi^-J/\psi$ with mass $3872\pm0.6\pm0.5$ MeV \cite{Choi:2003} and was confirmed by  CDF, D0 and BABAR Collaboration experiments \cite{Acosta:2003,Abazov:2004,Aubert:2004}. Several properties of the $X(3872)$ have been determined \cite{Choi:2011,Aaltonen:2009,Aubert:2008gu} and  CDF collaboration explained the X(3872) particle as a conventional charmonium $c\bar{c}$ state with $J^{PC}$ be either $1^{++}$ or $2^{-+}$ \cite{Abulencia:2006}. Recently BES III collaboration reported the first observation of process $e^-e^-{\rightarrow}\gamma X(3872)$ with mass $3871\pm0.7\pm0.2$ MeV\cite{Ablikim:2013c}. Barnes and Godfrey in 2003, evaluated the strong and electromagnetic decays and considered all possible 1D and 2P charmonium assignments for $X(3872)$ \cite{Barnes:2003}.

The $X(3915)$ was observed by S.K.Choi and his team at Belle Collaborations \cite{Choi:2004} and later on BABAR collaboration confirmed the existence of the charmonium-like resonance $X(3915)$ and measured its mass $3919.4\pm2.2\pm1.6$ MeV with the $J^{PC}=0^{++}$ option \cite{delAmoSanchez:2010,Lees:2012a}. This state is conventionally identified as the $\chi_{c0}(2P)$ charmonium \cite{Liu:2009,Zhou:2015}. The Belle Collaboration in the year 2005, observed the $Z(3930)$ resonance in the $\gamma\gamma\rightarrow D\bar{D}$ process \cite{Uehara:2005} with mass $3929\pm5\pm2$ MeV and considered {\bf it as} a strong candidate for the $\chi_{c2}$(2P) state. 
{\bf BABAR Collaboration was confirmed the  $Z(3930)$ resonance as the $\chi_{c2} (2P)$ state with mass $3926.7\pm2.7\pm1.1$ MeV and quantum numbers $J^{PC} = 2^{++}$  \cite{Aubert:2010}.}

In the year 2013, the BESIII collaboration observed a new structure with mass $3899\pm3.6\pm4.9$ MeV in the $\pi^{\pm}J/\psi$ mass spectrum (referred as $Z_{c}(3900)$) \cite{Ablikim:2013b} and simultaneously  Belle collaboration also observed a structure with mass $3894.5\pm6.6\pm4.5$ MeV in the $\pi^{\pm}J/\psi$ mass spectrum \cite{Liu:2013}. Observations of Xiao and his team, based on $e^+e^-$ annihilations at $\sqrt{s}=4170$ MeV, provide independent confirmation of the existence of the $Z_{c}^{\pm}(3900)$ state and provide new evidence for the existence of the neutral member $Z_{c}^{0}$(3900) \cite{Xiao:2013}.  Recently BES III Collaboration performed an analysis {\bf with favor to} the assignment of the $J^{P}=1^{+}$ quantum numbers \cite{Ablikim:2015b}.

In year 2009, CDF collaboration reported evidence for a narrow structure near $J/\psi\phi$ threshold in $B^+ {\rightarrow} J/\psi\phi K^+$ decays with mass $4143\pm2.9\pm1.2$ MeV \cite{Aaltonen:2009a} and recently observed by the CMS \cite{Chatrchyan:2013} and D0 \cite{Abazov:2015,Abazov:2013} collaborations.
It has been suggested that the $X(4140)$ resonance could be a molecular state \cite{Liu:2009a,Branz:2009,Albuquerque:2009,Ding:2009}, a tetra-quark state \cite{Stancu:2009,Wang:2015a,Anisovich:2015a} or a hybrid state \cite{Wang:2009,Mahajan:2009}. Searches for the narrow $X(4140)$ were negative in
LHCb \cite{Aaij:2012pz} and BaBar \cite{Lees2014a} experiments. In 2011, the CDF Collaboration observed the $X(4140)$ structure with a statistical significance greater than 5 standard deviations and also find evidence for a second structure  $X(4274)$ with a mass of $4274.4_{-6.7}^{+8.4}\pm1.9$ MeV \cite{Aaltonen:2011}. Very recently the LHCb Collaboration confirmed the resonance $X(4140)$ with mass $4146.5\pm4.5_{-2.8}^{+4.6}$ MeV and $X(4274)$ with mass $4273.3\pm8_{-3.6}^{+17.2}$ MeV in the $J/\psi\phi$ invariant mass distribution and determined their spin-parity quantum numbers to be  $J^{PC} = 1^{++}$ for both \cite{Aaij:2016nsc}. They also investigated two new structures named as the $X(4500)$ and $X(4700)$ in the high $J/\psi\phi$ mass region.  Ref.\cite{Lu:2016a} suggest that $X(4274)$ can be a good candidate  for the conventional $\chi_{c1} (3^{3}P_{1})$ state. Study of charmonium in relativistic Dirac formalism with linear confinement potential indicate that the $X(4140)$ state can be admixture of two P states whereas $X(4630)$ and $X(4660)$ are the admixed of S-D wave state\cite{Bhavsar:2018}.

Recently developed (GSPM) generalized screened potential model \cite{Gonzalez:2015}, the non-relativistic, Coulomb gauge QCD approach \cite{Guo:2014}, the light front quark model(LFQM) \cite{Ke:2013}, the relativistic quark model \cite{Ebert2002}, the effective field theory framework of potential non-relativistic QCD (pNRQCD) approach \cite{Brambilla2006}, the  effective
Lagrangian approach \cite{DeFazio:2008}, lattice
QCD \cite{Donald:2012ga,Liu:2012}, LCQCD and QCD sum rules \cite{Zhu:1998,Beilin:1984} and the widely used potential models \cite{Godfrey1985,Barnes:2005,Li:2012vc,Li:2009,Cao:2012,Segovia:2008,Eichten1978}, are different theoretical  model have been employed in theory to study the charmonium spectrum. The Cornell potential model is well known among the many phenomenologically successful potential models, which describes the charmonium system quite well.

The recent experimental results on new charmonium-like $X\,Y\,Z$  states indicate that they can be interpreted as above threshold charmonium levels and  cannot be assigned to any charmonium states in the conventional quark model. These experimental results,  motivate us and renewed theoretical interest to carry out a spectroscopic study and decay properties of charmonium.

In this article, to calculate the mass spectrum of the charmonium, we use Gaussian wave function both in position space as well as momentum space with a potential model, incorporating corrections to the kinetic energy of quarks as well as incorporating the relativistic correction of ${\cal{O}}\left(\frac{1}{m}\right)$ to the potential energy part of the Hamiltonian.  We  also investigate the Regge trajectories in both the $(M^{2}\rightarrow J)$ and $(M^{2}\rightarrow n)$ planes (where $J$ is the spin and $n$ is the principal quantum number) using our predicted masses for the charmonium, as the Regge trajectories play a significant role to identify the nature of current and future experimentally observed charmonium states. We also obtained the pseudoscalar and vector decay constants for charmonium as well as the radiative (Electric and Magnetic dipole)  transition rates and the annihilation decay. 

The article is organized as follows. Section {\ref{sec:mass}}, present the theoretical framework for the mass spectra, Section {\ref{sec:decay}} present the decay constants ($f_{P/V}$), Section {\ref{sec:E1M1}} present the radiative (E1 and M1) transitions in  and Section {\ref{sec:annihilation}} present annihilation decays. In Section {\ref{sec:Resu}}, we discuss results for the  mass spectra, ($f_{P/V}$) decays, E1 and M1 transition width as well as annihilation decays. The Regge trajectories from estimated masses in the $(J,M^{2})$ and $(n_r,M^{2})$ planes are in Section {\ref{sec:reg}}. Finally, we draw our conclusion in  Section-{\ref{sec:conclusion}}.

\section{Methodology}

\subsection{Cornell potential with ${\cal{O}}\left(\frac{1}{m}\right)$ corrections \label{sec:mass}}

Inspired by the extensive progress made in the experimental observation as well as the theoretical development of the charmonium, here we calculate the mass spectra and decay properties the charmonium within the widely used coulomb plus linear potential, Cornell potential \cite{Deng:2016,Godfrey1985,Barnes:2005,Godfrey:2015}. In this approach, we consider the relative corrections to the kinetic energy part and ${\cal{O}}\left(\frac{1}{m}\right)$ correction to the potential energy part \cite{Koma2006,Kher:2017,Kher:2017b}, which is inspired from the pNRQCD (potential non-relativistic quantum chromodynamics) \cite{Brambilla:2004jw,Brambilla2011,Brambilla:2014}. The  Cornell potential  working well for heavy light flavour, hence we employed it for heavy-heavy flavour.

We employ following Hamiltonian \cite{Gupta1995,Hwang1997,Kher:2017,Kher:2017b}  and quark-antiquark potential \cite{Koma2006} to study of the charmonium mass spectroscopy,

\begin{equation}
H=\sqrt{\mathbf{p}^{2}+m_{Q}^{2}}+\sqrt{\mathbf{p}^{2}+m_{\bar{Q}}^{2}}+V(\mathbf{r}),\label{Eq:hamiltonian}
\end{equation} 
\begin{equation}
V\left(r\right)=V^{\left(0\right)}\left(r\right)+\left(\frac{1}{m_{Q}}+\frac{1}{m_{\bar{Q}}}\right)V^{\left(1\right)}\left(r\right)+{\cal O}\left(\frac{1}{m_{}^{2}}\right).
\end{equation}
Here, $m_{Q}$($m_{\bar{Q}}$) is the quark(anti-quark) mass. and The Cornell-like potential $V^{\left(0\right)}$ \cite{Eichten1978} and leading order perturbation theory yields  $V^{\left(1\right)}\left(r\right)$ are, 
\begin{equation}\label{pote}
V^{\left(0\right)}(r)=-\frac{4\alpha_{S}\left({M^{2}}\right)}{3r}+Ar+V_{0}
\end{equation}
\begin{equation}
V^{\left(1\right)}\left(r\right)=-C_{F}C_{A}\alpha_{s}^{2}/4r^{2}
\end{equation}
where  
$\alpha_{S}\left({M^{2}}\right)$, $A$, $V_{0}$ and $C_{F}=4/3$, $C_{A}=3$ is the strong running coupling constant, potential parameter, potential constant and the Casimir charges respectively.
{\bf This correction was original studied by Y.Koma, where the relativistic correction to the QCD static potential ${\cal{O}}\left(\frac{1}{m}\right)$ was investigated non-perturbatively.  This correction is found to be similar to the Coulombic term of the static potential when applied to charmonium. The leading order corrections are classified in powers of the inverse of heavy quark mass}\cite{Koma2006}. 

Here, to estimate the expected values of the Hamiltonian with the Ritz variational strategy,  we use Gaussian wave function in position space {\bf as well as} in momentum space \cite{Kher:2017,Kher:2017b} has the form 
\begin{eqnarray}
R_{nl}(\mu,r) & = & \mu^{3/2}\left(\frac{2\left(n-1\right)!}{\Gamma\left(n+l+1/2\right)}\right)^{1/2}\left(\mu r\right)^{l}\times\nonumber \\
 &  & e^{-\mu^{2}r^{2}/2}L_{n-1}^{l+1/2}(\mu^{2}r^{2})
\end{eqnarray} and 
\begin{eqnarray}
R_{nl}(\mu,p) & = & \frac{\left(-1\right)^{n}}{\mu^{3/2}}\left(\frac{2\left(n-1\right)!}{\Gamma\left(n+l+1/2\right)}\right)^{1/2}\left(\frac{p}{\mu}\right)^{l}\times\nonumber \\
 &  & e^{-{p}^{2}/2\mu^{2}}L_{n-1}^{l+1/2}\left(\frac{p^{2}}{\mu^{2}}\right)
\end{eqnarray} respectively with the Laguerre polynomial $L$ and the variational parameter $\mu$. We estimated $\mu$ for each state, for the prefer value of $A$, using \cite{Hwang1997},
\begin{equation}
\left\langle{K.E.}\right\rangle =\frac{1}{2} \left\langle{\frac{rdV}{dr}}\right\rangle\label{Eq:virial theorem}
\end{equation} 

 To integrate relativistic correction, we enlarge Hamiltonian  Eq.(\ref{Eq:hamiltonian}) with powers up to  ${{\cal{O}}\left({\bf p}^{10}\right)}$ and ${\cal{O}}\left(\frac{1}{m}\right)$ at the kinetic energy and the potential energy part respectively \cite{Kher:2017}. {\bf We use a position space Gaussian wave-function to obtain expected value of the potential energy part whereas for the kinetic energy part, we use a momentum space wave-function using virial theorem Eq.(\ref{Eq:virial theorem}}).

We adapted the ground state center of weight mass and equated with the PDG data by fixing $A$, $\alpha_s$ and $V_0$  using the following equation \cite{Rai2008,Rai2002}:
\begin{equation}
M_{SA}=M_{P}+\frac{3}{4}(M_{V}-M_{P}),\label{Eq:rai1-1}
\end{equation}
We also forecast  the center of weight mass for the $nJ$ state as \cite{Rai2008}:
\begin{equation}
M_{CW,n}=\frac{\Sigma_{J}(2J+1)M_{nJ}}{\Sigma_{J}(2J+1)}\label{Eq:rai2-1}
\end{equation}

In the case of quarkonia, bound states are represented by $n^{2S+1}L_{J}$, identified with
the $J^{PC}$ values, with $\vec{J}=\vec{L}+\vec{S}$,  $\vec{S}=\vec{S}_{Q}+\vec{S}_{\bar{Q}}$, parity $P=(-1)^{L+1}$ and the charge conjugation $C=(-1)^{L+S}$ with $(n,L)$ being the radial quantum numbers. The spin dependent interaction are required to remove the degeneracy of charmonium states and can be written as \cite{Barnes:2005,Eichten:2008,Voloshin:2007,Lakhina2006}.
\begin{eqnarray}
V_{SD} & = & V_{LS}(r)\left(\vec{L}\cdot\vec{S}\right)+ V_{SS}(r)\left[S\left(S+1\right)-\frac{3}{2}\right]+\nonumber\\ 
 & &  V_{T}(r)\left[S\left(S+1\right)-\frac{3\left(\vec{S}\cdot\vec{r}\right)\left(\vec{S}\cdot\vec{r}\right)}{r^{2}}\right]
\end{eqnarray}
where the spin-spin, the spin-orbit and the tensor interactions can be written in terms of the vector and scalar parts of the $V(r)$ as by \cite{Voloshin:2007}
\begin{eqnarray}
V_{SS}(r) & = & \frac{1}{3m_{Q}^{2}}\nabla^{2}V_{V} =\frac{16\pi\alpha_{s}}{9m_{Q}^{2}}\delta^{3}\left(\vec{r}\right),
\end{eqnarray}
\begin{eqnarray}
V_{LS}(r) & = & \frac{1}{2m_{Q}^{2}r}\left(3\frac{dV_{V}}{dr}-\frac{dV_{S}}{dr}\right),
\end{eqnarray}
\begin{eqnarray}
V_{T}(r) & = & \frac{1}{6m_{Q}^{2}}\left(3\frac{d^{2}V_{V}}{dr^{2}}-\frac{1}{r}\frac{dV_{V}}{dr}\right),
\end{eqnarray}
\noindent where
$V_{V}(=-\frac{4\alpha_{s}}{3r})$ is the coulomb part and $V_{S}( = Ar )$ is the confining part of Eq.(\ref{pote})

In the present study, the quark masses is  $m_{c}=1.55$ ~GeV  to reproduce the ground state masses of the charmonium. The fitted potential parameters are $A=0.160\,\, GeV^{2}$, $\alpha_s=0.333$ and $V_0= -0.23074\,\, GeV$.

\subsection{Decay Constants ($f_{P/V}$)}
\label{sec:decay}
The decay constants with the QCD correction factor are computed using the Van-Royen-Weisskopf formula \cite{VanRoyen1967,Braaten1995},
\begin{equation}\label{Eq:decayconst}
f_{P/V}^{2}=\frac{12\left|\psi_{P/V}(0)\right|^{2}}{M_{P/V}}\left(1-\frac{\alpha_{S}}{\pi}\left[2-\frac{m_{Q}-m_{\bar{q}}}{m_{Q}+m_{\bar{q}}}\ln\frac{m_{Q}}{m_{\bar{q}}}\right]\right);\end{equation}
The Eq.(\ref{Eq:decayconst}) also gives the inequality\cite{Hwang1997a}
\begin{equation}\label{eq:ineq}
\sqrt{m_v}f_v \geq \sqrt{m_p}f_p
\end{equation}Our results are in accordance with Eq.(\ref{eq:ineq}) and tabulated in Table(\ref{tab:decaycc}). The value in parenthesis is the decay constant with QCD correction.

\subsection{Radiative Transitions \label{sec:E1M1}}
The radiative transition is influenced by the matrix element of the $EM$ current between the initial $i$ and final $f$ quarkonium state, i.e., $\langle f\mid j_{em}^{\mu}\mid i\rangle$. The electric dipole $(E1)$ or magnetic dipole $(M1)$ transition are leading order transition amplitudes \cite{Ding:2007,Lu2016,Guo:2010a}.

The E1 matrix elements are estimated by\cite{Radford2009}
\noindent 
\begin{eqnarray}
\Gamma_{(E1)}\left(n^{2S+1}L_{J}\rightarrow n^{'2S^{'}+1}L_{J^{'}}^{'}+\gamma\right)=\qquad&\qquad \qquad& \nonumber\\ 
\frac{4\alpha e_{Q}^{2}}{3} \frac{E_{\gamma}^{3}E_{f}}{M_{i}}C_{fi}\delta_{SS^{'}}\times\left|\left\langle f\left|r\right|i\right\rangle \right|^{2}
\end{eqnarray}
where Photon energy  ${\bf E_{\gamma}=\frac{M_{i}^{2}-M_{f}^{2}}{2M_{i}^{2}}}$;  the fine structure constant ${\bf \alpha=1/137}$; the quark charge $e_{Q}$ in units of the electron charge and the energy of final state  $E_f$. The angular momentum matrix element $C_{fi}$ is
\begin{equation}
C_{fi}=max\left(L,L^{'}\right)\left(2J^{'}+1\right)\left\{ \begin{array}{ccc}
L^{'} & J^{'} & S\\
J & L & 1
\end{array}\right\}^{2}
\end{equation}
where$\left\{ :::\right\} $ is a 6-j symbol. The matrix elements $ \langle n^{'2S^{'}+1}L_{J^{'}}^{'}\mid r\mid n^{2S+1}L_{J}\rangle $ were evaluated using the wave-functions 
\begin{equation}
\left\langle f \left| r \right| i \right\rangle=\int dr R_{n_{i}l_{i}}\left(r\right)R_{n_{f}l_{f}}\left(R\right)
\end{equation}

The M1 radiative transitions are evaluated {\bf using} the following expression \cite{Segovia:2016,Barnes:2005}
\begin{equation}\label{eq:m1}
 \Gamma_{M1}\left(n^{2S+1}L_{J}\rightarrow n^{'2S^{'}+1}L_{J^{'}}^{'}\right)=\frac{4\alpha e_{Q}^{2}}{3m_{Q}^2} \frac{E_{\gamma}^{3}E_{f}}{M_{i}}S_{fi}\left|\mathcal{M}_{fi} \right|^{2},\end{equation} 
where, 
\begin{equation}
\mathcal{M}_{fi} =\int dr R_{n_{i}l_{i}}\left(r\right)j_{0}\left(E_{\gamma}r/2\right)R_{n_{f}l_{f}}\left(R\right)
\end{equation} and
\begin{eqnarray}
S_{fi}&=& 6\left(2S+1\right)\left(2S^{'}+1\right)\left(2J^{'}+1\right)\times \nonumber \\
& & \left\{ \begin{array}{ccc}
 J & 1 &  J^{'}\\
 S^{'} & L & S
\end{array}\right\}^{2}\left 
 \{ \begin{array}{ccc}
 1 & 1/2 &  1/2\\
 1/2 & S^{'} & S
\end{array}\right\}^{2}
\end{eqnarray}
here L = 0 for S-waves and $j_{0}(x)$ is the spherical Bessel function.

The E1 and M1 radiative transition widths are listed in table (\ref{tab:E1CC}) and (\ref{tab:M1CC}) respectively.

\subsection{Annihilation Decays \label{sec:annihilation}}
Decays of quarkonia states into leptons or photons or gluons is extremely useful for the production and identification of resonances as well as the leptonic decay rates of quarkonia. {\bf It} can also assist to recognize conventional mesons and multi-quark structures \cite{Kwong:1987,Kwong:1988}.

\subsubsection{Leptonic decays}
The $^{3}S_{1}$ and $^{3}D_{1}$ states have $J^{PC}=1^{--}$ quantum numbers, annihilate into lepton pairs through a single virtual photon. The leptonic decay width of the ($^{3}S_{1}$) and ($^{3}D_{1}$) states of charmonium  including first order radiative QCD correction is given by \cite{Segovia:2016,Kwong:1987,Bradley:1980}.
\begin{equation}
\varGamma\left(n^{3}S_{1}\rightarrow e^{+}e^{-}\right)=\frac{4e_{Q}^{4}\alpha^{2}\mid R_{nS}\left(0\right)\mid^{2}}{M_{nS}^{2}}\left(1-\frac{16\alpha_{s}}{3\pi}\right)
\end{equation}
\begin{equation}
\varGamma\left(n^{3}D_{1}\rightarrow e^{+}e^{-}\right)=\frac{25e_{Q}^{2}\alpha^{2}\mid R_{nD}^{\prime\prime}\left(0\right)\mid^{2}}{2m_{Q}^{4}M_{nD}^{2}}\left(1-\frac{16\alpha_{s}}{3\pi}\right)
\end{equation}
where, 
$M_{nS}$ is mass of the decaying charmonium state.

\subsubsection{Decay into photons }
The annihilation decay of the charmonium states into two or three photon, without and/or with radiative QCD corrections are given by\cite{Segovia:2016,Kwong:1987}
\begin{equation}
\varGamma\left(n^{1}S_{0}\rightarrow\gamma\gamma\right)=\frac{3e_{Q}^{4}\alpha^{2}\mid R_{nS}\left(0\right)\mid^{2}}{m_{Q}^{2}}\left(1-\frac{3.4\alpha_{s}}{\pi}\right)
\end{equation}
\begin{equation}
\varGamma\left(n^{3}P_{0}\rightarrow\gamma\gamma\right)=\frac{27e_{Q}^{4}\alpha^{2}\mid R_{nP}^{\prime}\left(0\right)\mid^{2}}{m_{Q}^{4}}\left(1+\frac{0.2\alpha_{s}}{\pi}\right)
\end{equation}
\begin{equation}
\varGamma\left(n^{3}P_{2}\rightarrow\gamma\gamma\right)=\frac{36e_{Q}^{4}\alpha^{2}\mid R_{nP}^{\prime}\left(0\right)\mid^{2}}{5m_{Q}^{4}}\left(1-\frac{16\alpha_{s}}{3\pi}\right)
\end{equation}
\begin{eqnarray}
\varGamma\left(n^{3}S_{1}\rightarrow3\gamma\right)&=&\frac{4 (\pi^{2}-9)e_{Q}^{6}\alpha^{3}\mid R_{nS}\left(0\right)\mid^{2}}{3\pi m_{Q}^{2}} \nonumber \times\\ &&\left(1-\frac{12.6\alpha_{s}}{\pi}\right)
\end{eqnarray}

\subsubsection{Decay into gluons }
The annihilation decay of the charmonium states into two or three gluon as well as  into gluons with photon and light quark, without and/or with radiative QCD correction are given by\cite{Segovia:2016,Kwong:1987,Kwong:1988,Belanger:1987}
\begin{equation}
\varGamma\left(n^{1}S_{0}\rightarrow gg\right)=\frac{2\alpha_{s}^{2}\mid R_{nS}\left(0\right)\mid^{2}}{3m_{Q}^{2}}\left(1+\frac{4.8\alpha_{s}}{\pi}\right)
\end{equation}
\begin{equation}
\varGamma\left(n^{3}P_{0}\rightarrow gg\right)=\frac{6\alpha_{s}^{2}\mid R_{nP}^{\prime}
\left(0\right)\mid^{2}}{m_{Q}^{4}}
\end{equation}
\begin{equation}
\varGamma\left(n^{3}P_{2}\rightarrow gg\right)=\frac{8\alpha_{s}^{2}\mid R_{nP}^{\prime}
\left(0\right)\mid^{2}}{5m_{Q}^{4}}
\end{equation}
\begin{equation}
\varGamma\left(n^{1}D_{2}\rightarrow gg\right)=\frac{2\alpha_{s}^{2}\mid R_{nD}^{\prime\prime}
\left(0\right)\mid^{2}}{3\pi m_{Q}^{6}}
\end{equation}
\begin{eqnarray}
\varGamma\left(n^{3}S_{1}\rightarrow3g\right)&=&\frac{10 (\pi^{2}-9)\alpha_{s}^{3}\mid R_{nS}\left(0\right)\mid^{2}}{81\pi m_{Q}^{2}}\nonumber \times\\ &&\left(1-\frac{3.7\alpha_{s}}{\pi}\right)
\end{eqnarray}
\begin{equation}
\varGamma\left(n^{1}P_{1}\rightarrow3g\right)=\frac{20 \alpha_{s}^{3}\mid R_{nP}^{\prime}\left(0\right)\mid^{2}}{9\pi m_{Q}^{4}}ln(m_{Q}\langle r \rangle)
\end{equation}
\begin{equation}
\varGamma\left(n^{3}D_{1}\rightarrow3g\right)=\frac{760 \alpha_{s}^{3}\mid R_{nP}^{\prime\prime}\left(0\right)\mid^{2}}{81\pi m_{Q}^{6}}ln(4m_{Q}\langle r \rangle)
\end{equation}
\begin{equation}
\varGamma\left(n^{3}D_{2}\rightarrow3g\right)=\frac{10 \alpha_{s}^{3}\mid R_{nP}^{\prime\prime}\left(0\right)\mid^{2}}{9\pi m_{Q}^{4}}ln(4m_{Q}\langle r \rangle)
\end{equation}
\begin{equation}
\varGamma\left(n^{3}D_{3}\rightarrow3g\right)=\frac{40 \alpha_{s}^{3}\mid R_{nP}^{\prime\prime}\left(0\right)\mid^{2}}{9\pi m_{Q}^{6}}ln(4m_{Q}\langle r \rangle)
\end{equation}
\begin{eqnarray}
\varGamma\left(n^{3}S_{1}\rightarrow \gamma gg\right)&=&\frac{8 (\pi^{2}-9)e_{Q}^{2}\alpha\alpha_{s}^{2}\mid R_{nS}\left(0\right)\mid^{2}}{9\pi m_{Q}^{2}}\nonumber \times\\
&&\left(1-\frac{6.7\alpha_{s}}{\pi}\right)
\end{eqnarray}
\begin{equation}
\varGamma\left(n^{3}P_{1}\rightarrow q\bar{q}+g\right)=\frac{8\eta_{f} \alpha_{s}^{3}\mid R_{nP}^{\prime}\left(0\right)\mid^{2}}{9\pi m_{Q}^{4}}ln(m_{Q}\langle r \rangle)
\end{equation}

 The calculated annihilation decay width of charmonium are listed in Tables(\ref{tab:annihi2e} \text{to} \ref{tab:annihiqqg}).

\end{multicols}

    \begin{figure}
      \centering
     \includegraphics[bb=30bp 60bp 750bp 550bp,clip,width=0.80\textwidth]{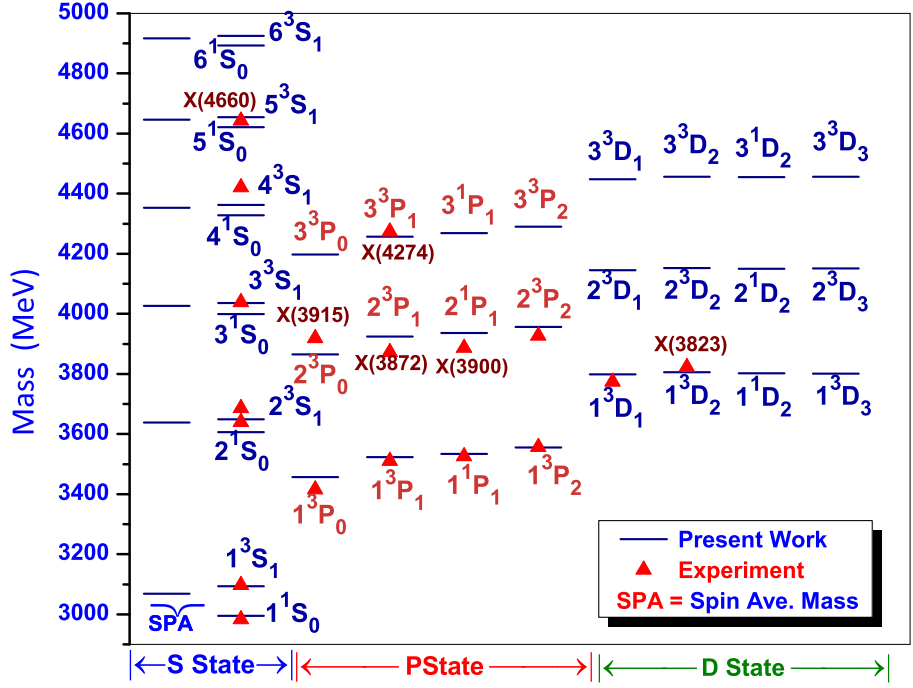}
     \caption{Mass spectrum.\label{fig:MassCC}}
     \end{figure}

  \begin{table}
  \caption{Pseudoscalar and vector decay constants (in ${GeV}$).\label{tab:decaycc} }
  \noindent \centering{}%
   \begin{tabular}{lllllll}
  \hline
  Decay& State & Our Work & Expt.\cite{PDGlatest} & \cite{Bhaghyesh:2011}  & \cite{Negash:2015}&\cite{Bhavsar:2018} \\
  \addlinespace[3pt]
  \hline
  \addlinespace[5pt]
  \(f_P\) & 1S & 0.501(0.395) &  $\bf 0.335\pm0.075$ & 0.471(0.360) & 0.404 &  \\
  & 2S & 0.301(0.237) &  &0.344(0.286) & 0.331&\\
  & 3S & 0.264(0.208) &  &0.332(0.254) & 0.291&\\
  & 4S & 0.245(0.193) &  & 0.312(0.239)&& \\
  & 5S & 0.233(0.184) &  & & &\\
  & 6S & 0.224(0.177) &  & && \\
  \hline 
  \addlinespace[5pt]
  \(f_V\)&1S & 0.510(0.402) & $\bf 0.411\pm0.005$ & 0.462(0.317) & 0.375& 0.420  \\
  &2S & 0.303(0.239) & $\bf 0.271\pm0.008$ &0.369(0.253) &0.295&0.285 \\
  &3S & 0.265(0.209) & $\bf 0.174\pm0.018$ &0.329(0.226) &0.261 &0.218\\
  &4S & 0.240(0.194) &  &0.310(0.212) &0.240&0.166 \\
  &5S & 0.234(0.185) &  & 0.290(0.199)&&0.106 \\
  &6S & 0.225(0.177) &  & & &\\
  \hline
  \end{tabular}
   \end{table}

\begin{table}
\caption{S-P-D-wave center of weight masses (in GeV). (LP = Linear potential model, SP = Screened potential model, NR = Non-relativistic and  RE = Relativistic)\label{tab:ccmsa}}
\scalebox{0.90}{
 \begin{tabular}{cccccccccccccc}
 \hline
  & \multicolumn{2}{c}{This work} &  & \multicolumn{10}{c}{Others Theory $M_{SA}$ in (GeV) }\\
$nL$ & $\mu$  & $M_{SA}$ & Expt.\cite{PDGlatest} &  \multirow{2}{*}{LP (SP) \cite{Deng:2016}} & \multirow{2}{*}{\cite{Yang:2015cc}} & \multirow{2}{*}{\cite{Ebert2011}} & \multirow{2}{*}{\cite{Cao:2012}}& \multirow{2}{*}{NR (GI)\cite{Barnes:2005}} & \multirow{2}{*}{\cite{Li:2009}} & \multirow{2}{*}{\cite{Radford:2007}}
&\multirow{2}{*}{\cite{Ebert:2002}}
&\multirow{2}{*}{RE(NR)\cite{Sultan:2014}} &\multirow{2}{*}{\cite{Godfrey:1985}}\\
&$(GeV)$ & $(GeV)$&$(GeV)$ & & &  & & & & & & \\ 
 \hline
 \addlinespace[3pt] 
$1S$ & 0.716  & 3.068 & 3.068 & 3.068 (3.069)& 3.090 & 3.067& 3.061& 3.063 (3.067) & 3.068 &3.068 &  3.068&3.068 (3.063) &3.068 \\
$2S$ &  0.469  & 3.638 & 3.674 & 3.668 (3.668)& 3.667 & 3.673& 3.676 & 3.662 (3.663)& 3.661 &3.664 &  3.662& 3.657 (3.661)& 3.665\\
$3S$ & 0.412  & 4.027 & & 4.071 (4.024)& 4.070 & 4.027&  4.080 & 4.065 (4.091) & 4.014 &4.075 &   4.064& 4.051 (4.064)& 4.090\\
$4S$ & 0.382  & 4.353 &  & 4.406 (4.277)& 4.408 & 4.421& 4.406 & 4.400 (4.444)& 4.267 & & & 4.350 (4.400) & \\
$5S$ & 0.363  & 4.646 &  & 4.706 (4.469)& 4.710 & 4.831& &  & 4.459 & & & 4.655 (4.694) & \\
$6S$ & 0.349  & 4.917 &  &  & 4.987 & 5.164& &  & 4.603 & & & 4.907 (4.973)& \\
 \addlinespace[5pt]   
$1P$ & 0.484  & 3.534 & 3.525 & 3.524 (3.527) & 3.523 & 3.525&3.525& 3.522 (3.523) & 3.524 & 3.526&  3.526& 3.554 (3.519)&3.523 \\
$2P$ & 0.416  & 3.936 &  & 3.945 (3.919)& 3.941 & 3.926&3.945 & 3.942 (3.961) & 3.913 &3.960 &  3.945 & 3.963 (3.938)& 3.962 \\
$3P$ & 0.384  & 4.269 &  & 4.291 (4.238)& 4.289 & 4.337& 4.316 & 4.286 (4.323) & 4.188 & &  & 4.296 (4.283)& \\
 \addlinespace[5pt]     
$1D$ & 0.437  & 3.802 &  & 3.805 (3.805) & 3.798 & 3.803& 3.815 & 3.800 (3.849) & 3.796 & 3.823 &   3.811& 3.839 (3.799)&3.837\\
$2D$ & 0.396  & 4.150 &  & 4.164 (4.108)& 4.160 & 4.196& 4.165& 4.159 (4.209)& 4.099 &4.190 &  & 4.187 (4.158) & 4.210\\
$3D$ & 0.372  & 4.455 &  & 4.478 (4.336) & 4.478 & 4.455&4.522 &  & 4.327 & & & 4.486 (4.473)& \\
\hline
\end{tabular}}
\end{table}

\begin{table}
\caption{Hyperfine and fine splittings(in MeV). (LP = Linear potential model, SP = Screened potential model, NR = Non-relativistic and  RE = Relativistic)\label{tab:masssplit}}
\scalebox{0.93}{
 \begin{tabular}{lllllllllllll}
 \hline
 \multirow{2}{*}{Splitting}   & This & Expt. &  \multicolumn{10}{c}{Others}\\
  &   work & \cite{PDGlatest} &  \cite{Deng:2016} & \cite{Yang:2015cc} & \cite{Ebert2011} & \cite{Barnes:2005} & \cite{Cao:2012}& \cite{Li:2009} &\cite{Radford:2007} & \cite{Sultan:2014}&\cite{Ebert:2002}&\cite{Bhavsar:2018} \\
   &    &  &  LP(SP) &  &  & NR(GI) & &  & &  RE (NR)&& \\
  \hline
   \addlinespace[3pt]
 
m($1^{3}S_{1}$)-m($1^{1}S_{0}$)   & 99 & $113.3\pm0.7$ & 114 (113)& 116 & 115 & 108 (123) & 100& 118 & 117 & 102 (108) & 117&119\\
  
m($2^{3}S_{1}$)-m($2^{1}S_{0}$) & 43 & $46.7\pm1.3$ & 44  (42)& 11 & 51 & 42 (53) &38 & 50 & 89& 33 (42)& 98& 54\\
  
m($3^{3}S_{1}$)-m($3^{1}S_{0}$)   & 36 &  & 30 (26)& 9 &50 & 29 (36) &29 &  31 &81 & 30 (29) &97& 32\\

m($4^{3}S_{1}$)-m($4^{1}S_{0}$) & 34 &  & 24 (17)& 6 & 26 & 22 (25)& 20 & 23 && 24 (22) & & 4.3 \\
  
m($5^{3}S_{1}$)-m($5^{1}S_{0}$) & 32 &  & 21 (13) & 6 & 26 & & & 17 & & 22 (19) & &2.3 \\

m($6^{3}S_{1}$)-m($6^{1}S_{0}$) & 32 &  &  & 5 & 12 & & & 10 & & 19 (17)& & \\
 \addlinespace[5pt]
m($1^{3}P_{2}$)-m($1^{3}P_{1}$)   & 33 & $45.5\pm0.2$  & 36 (32)& 47 & 44 & 51 (40) & 41& 44 & 50& 41 (44) & 46&\\

m($1^{3}P_{1}$)-m($1^{3}P_{0}$)  & 66 & $95.9\pm0.4$ & 101 (106) & 63 & 102 & 81 (65)& 52& 77 & 92& 71 (80) & 86&\\
 
m($2^{3}P_{2}$)-m($2^{3}P_{1}$)  & 31 &  & 30 (23) & 46 & 45 & 47 (26) & 38 &36& 54& 40 (40) &43&\\

m($2^{3}P_{1}$)-m($2^{3}P_{0}$)   & 59 & & 68 (66)& 59 & 36 & 73 (37)& 92 & 59 & 96& 66 (73) & 75&\\
 
 m($3^{3}P_{2}$)-m($3^{3}P_{1}$)  & 33 &  & 26  (19) & 44 & 35 & 46 (20)& 53 & 30 & & 45 (38) &&\\

 m($3^{3}P_{1}$)-m($3^{3}P_{0}$)  & 60 &  & 54 (46) & 58 &18 & 69 (25) & 81& 47 & & 63 (69) && \\
 \hline
\end{tabular}}
 \end{table}

  \begin{table*}
  \caption{Complete mass spectra (in GeV). (LP = Linear potential model, SP = Screened potential model, NR = Non-relativistic and  RE = Relativistic, )\label{tab:massescc}}
 \scalebox{0.90}{ 
   \begin{tabular}{llllllllllllll}
   \hline
   State & \multirow{2}{*}{$J^{P}$} & This & Expt. &  \multicolumn{10}{c}{Others}\\
   $n^{2S+1}L_{J}$ &  & work & \cite{PDGlatest} &  LP (SP) \cite{Deng:2016} & \cite{Yang:2015cc} & \cite{Ebert2010} & NR (GI) \cite{Barnes:2005} & \cite{Cao:2012}& \cite{Li:2009} &\cite{Radford:2007} &\cite{Ebert:2002} &RE (NR)\cite{Sultan:2014} &\cite{Godfrey:1985}\\
    \hline
     \addlinespace[3pt]
  $1^{1}S_{0}$ & $0^{-+}$ & 2.995 & 2.984 & 2.983 (2.984) & 3.069 & 2.981 & 2.982 (2.975) & 2.978 & 2.979& 2.980 & 2.979& 2.992 (2.982)& 2.97 \\
  $1^{3}S_{1}$ & $1^{--}$ & 3.094 & 3.097 & 3.097 (3.097)& 3.097 & 3.096 & 3.090 (3.098) & 3.088& 3.097 & 3.097 &  3.096& 3.094 (3.090)&3.10\\
  $2^{1}S_{0}$ & $0^{-+}$ & 3.606 & 3.639 & 3.635 (3.637) & 3.659 & 3.635 & 3.630 (3.623) & 3.647& 3.623 & 3.597&  3.588& 3.625 (3.630) &3.62 \\
  $2^{3}S_{1}$ & $1^{--}$ & 3.649 & 3.686 & 3.679 (3.679)& 3.670 & 3.686 & 3.672 (3.676) &3.685 & 3.673 & 3.686&   3.686& 3.668 (3.672) & 3.68 \\
  $3^{1}S_{0}$ & $0^{-+}$ & 4.000 & & 4.048 (4.004) & 4.063 & 3.989 & 4.043 (4.064)&4.058 & 3.991 & 4.014&   3.991 &4.029 (4.043) &4.06 \\
  $3^{3}S_{1}$ & $1^{--}$ & 4.036 & 4.039 & 4.078 (4.030)& 4.072 & 4.039 & 4.072 (4.100) &4.087 &  4.022 &4.095 &  4.088 &4.059 (4.072) &4.10\\
  $4^{1}S_{0}$ & $0^{-+}$ & 4.328 &  & 4.388 (4.264)& 4.403 & 4.401 & 4.384 (4.425)& 4.391 & 4.250 & &   & 4.332 (4.388)& \\
  $4^{3}S_{1}$ & $1^{--}$ & 4.362 & 4.421 & 4.412 (4.281)& 4.409 & 4.427 & 4.406 (4.450) & 4.411 & 4.273 & 4.433&   & 4.356 (4.406)&4.45 \\
  $5^{1}S_{0}$ & $0^{-+}$ & 4.622 &  & 4.690 (4.459) & 4.705 & 4.811 & & & 4.446 & &  &4.639 (4.685)& \\
  $5^{3}S_{1}$ & $1^{--}$ & 4.654 & 4.643 & 4.711 (4.472)& 4.711 & 4.837 & & & 4.463 & &  &4.661 (4.704)& \\
  $6^{1}S_{0}$ & $0^{-+}$ & 4.893 &  &  & 4.983 & 5.155 &  & & 4.595 & &  &4.893 (4.960)& \\
  $6^{3}S_{1}$ & $1^{--}$ & 4.925 &  &  & 4.988 & 5.167 & & & 4.605 & &  &4.912 (4.977)& \\
   \addlinespace[5pt]
   $1^{3}P_{0}$ & $0^{++}$ & 3.457 & 3.415 & 3.415 (3.415) & 3.440 & 3.413 & 3.424 (3.445) & 3.366& 3.433 & 3.416&   3.424&3.472 (3.424) &3.44\\
    $1^{3}P_{1}$ & $1^{++}$ & 3.523 & 3.511 & 3.516 (3.521) & 3.503 & 3.511 & 3.505 (3.510) & 3.518 & 3.510 & 3.508& 3.510 & 3.543 (3.505) &3.51\\
    $1^{1}P_{1}$ & $1^{+-}$ & 3.534 & 3.525 & 3.522 (3.526) & 3.526 & 3.525 & 3.516 (3.517) & 3.527 & 3.519 & 3.527&  3.526 & 3.544 (3.516) &3.52\\
    $1^{3}P_{2}$ & $2^{++}$ & 3.556 & 3.556 & 3.552 (3.553)& 3.550 & 3.555 & 3.556 (3.550)& 3.559& 3.554 & 3.558&   3.556&3.584 (3.549) &3.55\\
    $2^{3}P_{0}$ & $0^{++}$ & 3.866 & 3.918 & 3.869 (3.848)& 3.862 & 3.870 & 3.852 (3.916) & 3.843 & 3.842 & 3.844&   3.854& 3.885 (3.852)&3.92\\
    $2^{3}P_{1}$ & $1^{++}$ & 3.925 & 3.872 & 3.937 (3.914)& 3.921 & 3.906 & 3.925 (3.953)&3.935 & 3.901 & 3.940&  3.929& 3.951 (3.925)&3.95\\
    $2^{1}P_{1}$ & $1^{+-}$ & 3.936 &3.887 & 3.940 (3.916)& 3.944 & 3.926 & 3.934 (3.956) &3.942 & 3.908 &3.961 &  3.945 &3.951 (3.934) &3.96\\
    $2^{3}P_{2}$ & $2^{++}$ & 3.956 & 3.927 & 3.967 (3.937) & 3.967 & 3.949 & 3.972 (3.979) &3.973 & 3.937 & 3.994&  3.972 &3.994 (3.965) &3.98\\
    $3^{3}P_{0}$ & $0^{++}$ & 4.197 &  & 4.230 (4.146) & 4.212 & 4.301 &4.202 (4.292) & 4.208& 4.131 & &   & 4.219 (4.202)&\\
    $3^{3}P_{1}$ & $1^{++}$ & 4.257 & 4.273 & 4.284 (4.192) & 4.270 & 4.319 & 4.271 (4.317) & 4.299 & 4.178 & &   &4.283 (4.271) & \\
   
    $3^{1}P_{1}$ & $1^{+-}$ & 4.269 &  & 4.285 (4.193) & 4.292 & 4.337 & 4.279 (4.318) & 4.310& 4.184 & &   & 4.283 (4.279)&\\
    $3^{3}P_{2}$ & $2^{++}$ & 4.290 &  & 4.310 (4.311) & 4.314 & 4.354 & 4.317 (4.337)& 4.352 & 4.208 & &   & 4.328 (4.309)& \\
  
   \addlinespace[5pt]
   $1^{3}D_{1}$ & $1^{--}$ & 3.799 & 3.773 & 3.787 (3.792)& 3.759 & 3.783 & 3.785 (3.819) & 3.809& 3.787 & 3.804&   3.798& 3.830 (3.785)&3.82 \\
   $1^{3}D_{2}$ & $2^{--}$ & 3.805 & 3.822 & 3.807 (3.807) & 3.787 & 3.795 & 3.800 (3.838) & 3.820 & 3.798 & 3.824&   3.813&3.841 (3.800) &3.84\\
   $1^{1}D_{2}$ & $2^{-+}$ & 3.802 &  & 3.806 (3.805) & 3.799 & 3.807 & 3.799 (3.879) &3.815 & 3.796 & 3.824&  3.811& 3.837 (3.799)&3.84\\
   $1^{3}D_{3}$ & $3^{--}$ & 3.801 &  & 3.811 (3.808)& 3.823 & 3.813 & 3.806 (3.849) &3.813 & 3.799 & 3.831 &  3.815&3.844 (3.805) &3.84\\
   $2^{3}D_{1}$ & $1^{--}$ & 4.145 & 4.191 & 4.144 (4.095)& 4.119 & 4.150 & 4.142 (4.194) &4.154 & 4.089 &4.164 &  & 4.174 (4.141)&4.19\\
   $2^{3}D_{2}$ & $2^{--}$ & 4.152 &  & 4.165 (4.109) & 4.148 & 4.190 & 4.158 (4.208) &4.169 &4.100 &4.189 &   & 4.187 (4.158)&4.21\\
   $2^{1}D_{2}$ & $2^{-+}$ & 4.150 &  & 4.164 (4.108)& 4.160 & 4.196 & 4.158 (4.208) & 4.165& 4.099 & 4.191& & 4.183 (4.158)&4.21\\
   $2^{3}D_{3}$ & $3^{--}$ & 4.151 &  & 4.172 (4.112)& 4.185 & 4.220 & 4.167 (4.217) & 4.166& 4.103 & 4.202&  & 4.195 (4.165)&4.22\\
   $3^{3}D_{1}$ & $1^{--}$ & 4.448 &  & 4.456 (4.324)& 4.437 & 4.448 &  & 4.502 & 4.317 & 4.477&  & 4.470 (4.455) &4.52 \\
   $3^{3}D_{2}$ & $2^{--}$ & 4.456 &  & 4.478 (4.337)& 4.466 & 4.456 &  &4.524 & 4.327 & & & 4.485 (4.472)& \\
   $3^{1}D_{2}$ & $2^{-+}$ & 4.455 &  & 4.478 (4.336)& 4.478 & 4.455 &  & 4.524 & 4.326 & & & 4.480 (4.472)& \\
   $3^{3}D_{3}$ & $3^{--}$ & 4.457 &  & 4.486 (4.340) & 4.503 & 4.457 &  & 4.527 & 4.331 & & & 4.497 (4.481)& \\
  \hline
  \end{tabular}}
  \end{table*}

 \begin{table*}
 \caption{Electric dipole (E1) transitions widths of $c\overline{c}$ mesons. (LP = Linear potential model, SP = Screened potential model, NR = Non-relativistic and  RE = Relativistic, Here $E_{\gamma}$ in MeV and $\varGamma$ in KeV) \label{tab:E1CC} }
 \scalebox{0.90}{
 \begin{tabular}{ccccccccccccccc}
 \hline
 \addlinespace[3pt]
 \multicolumn{2}{c}{Transition} & \multicolumn{2}{c}{This work} & Expt.\cite{PDGlatest}& \multicolumn{10}{c}{Other work} \\
 Initial & Final & $E_{\gamma}$ & $\varGamma$ & $\varGamma$ & \cite{Li:2009} & \cite{Ebert:2002} & \cite{Parmar2010} & \cite{Barnes:2005}& \cite{Brambilla:2004}& \cite{Eichten:2002}& \cite{Segovia:2008}&\cite{Cao:2012}& \cite{Deng:2016}& \cite{Sultan:2014}\\
  &  &  &  &  &  &  &  & NR(GI)& & & & & LP(SP)& RE(NR)\\
 \hline
 \addlinespace[5pt]
 $1^{3}P_{2}$ & $1^{3}S_{1}$ & 432.31 & 233.85 & $406\pm31$ & 309 & 327 & 383& 424 (313) & 315& 315& & 405 & 327(338)& 437.5(424.5)\\
 $1^{3}P_{1}$ & $1^{3}S_{1}$ & 402.92 & 189.86 & $320\pm25$ & 244 & 265 & 361& 314 (239) & 241& 242&&341 &269 (278) & 329.5(319.5)\\
 $1^{1}P_{1}$ & $1^{1}S_{0}$ & 497.67 & 357.83 &  & 323 & 560 & 671& 498 (352) & 482& 482&&473 &361 (373)& 570.5(490.3)\\
 $1^{3}P_{0}$ & $1^{3}S_{1}$ & 344.13 & 118.29 & $131\pm14$ & 117 & 121 & 264 & 152  (114) & 120& 120& & 104 & 141(146)& 159.2(154.5)\\
 \hline
 \addlinespace[5pt]
 $2^{3}S_{1}$ & $1^{3}P_{2}$ & 91.58 & 7.07 & $26\pm1.5$ & 34 & 18.2 & & 38 (24) & 30.1& 29&28.6&39 & 36(44)& 35.5 (37.9) \\
 $2^{3}S_{1}$ & $1^{3}P_{1}$ & 123.46 & 10.39 & $27.9\pm1.5$ & 36 & 22.9 & & 54 (29) & 42.8& 41&33.0&38 & 45(48)&50.9 (54.2) \\
 $2^{3}S_{1}$ & $1^{1}P_{1}$ & 112.88 & 7.94 &  & 104 &  &  & && && & & \\
 $2^{3}S_{1}$ & $1^{3}P_{0}$ & 186.43 & 11.93 & $29.8\pm1.5$ & 25 & 26.3 & &63 (26) & 47& 46 &28.8&29 & 27(26)& 58.8 (62.6)\\
 $2^{1}S_{0}$ & $1^{3}P_{1}$ & 82.19 & 9.20 &  & & &  &  &&&& & & \\
 $2^{1}S_{0}$ & $1^{1}P_{1}$ & 71.49 & 6.05 &  &  & 6.2 & &49 (36)& 35.1& 35.1&& 56 &49 (52)& 45.2 (49.9)\\
 
   \hline
   \addlinespace[5pt]
 $1^{3}D_{3}$ & $1^{3}P_{2}$ & 237.31 & 237.51 &  & 323 & 156 & 432 & 272 (296)& 402&&& 302 & &397.7(271.1) \\
 $1^{3}D_{2}$ & $1^{3}P_{2}$ & 241.19 & 62.34 &  & 55 & 59 & 131& 64 (66)& 69.5& 56&& 82& 79(82)&96.52(64.06) \\
 $1^{3}D_{2}$ & $1^{3}P_{1}$ & 271.75 & 89.18 &  & 208 & 215 & 423 &307 (268)& 313& 260&&301& 281(291)& 438.2(311.2) \\
 $1^{3}D_{1}$ & $1^{3}P_{2}$ & 235.48 & 6.45 & $<21$ & 4.6 & 6.9 & 15.2 & 4.9 (3.3)& 3.88& 3.7& 3.3& 8.1&5.4 (5.7)& 4.73(4.86) \\
 $1^{3}D_{1}$ & $1^{3}P_{1}$ & 266.10 & 139.52 & $70\pm17$ & 93 & 135 & 246&125 (77) & 99& 94& 89.7&153& 115 (111)& 122.8(126.2) \\
 $1^{3}D_{1}$ & $1^{3}P_{0}$ & 326.57 & 343.87 & $172\pm30$ & 197 & 355 & 448 & 403 (213)& 299& 287& 221.7& 362&243 (232)& 394.6(405.4)\\
 \hline
   \addlinespace[5pt]
 $2^{3}P_{2}$ & $2^{3}S_{1}$ & 295.70 & 281.93 &  & 100 &  & 164 & 304 (207)&&& &264 & & 377.1(287.5)\\
 $2^{3}P_{1}$ & $2^{3}S_{1}$ & 266.71 & 206.87 &  & 60 &  & 174 &183 (183) &&&& 234& & 246.0(185.3)\\
 $2^{1}P_{1}$ & $2^{1}S_{0}$ & 315.84 & 343.55 &  & 108 &  & 333&280 (218) & &&&274& & 349.8(272.9) \\
 $2^{3}P_{0}$ & $2^{3}S_{1}$ & 210.86 & 102.23 &  & 44 &  & 112&64 (135) & &&&83& & 108.3(65.3)\\
 \hline
   \addlinespace[5pt]
 $2^{3}P_{2}$ & $1^{3}D_{3}$ & 152.16 & 33.27 &  &  &  &  & 88 (29)&&&& 76& & 60.67(78.69)\\
 $2^{3}P_{2}$ & $1^{3}D_{2}$ & 148.18 & 5.49 &  &  &  &  &17 (5.6)& &&& 10& & 11.48(15.34)\\
 $2^{3}P_{2}$ & $1^{1}D_{2}$ & 151.21 & 5.83 &  &  &  &  & & &&&& &  \\
 $2^{3}P_{2}$ & $1^{3}D_{1}$ & 154.03 & 0.41 &  &  &  &  &1.9 (1.0)& &&& 0.64& & 2.31(1.67)\\
 $2^{3}P_{1}$ & $1^{3}D_{1}$ & 123.91 & 5.35 &  &  &  &  & 22 (21)&&&& 11& & 31.15(21.53)\\
 $2^{3}P_{0}$ & $1^{3}D_{1}$ & 65.87 & 3.21 &  &  &  &  &13 (51)&& && 1.4& &33.24(13.55) \\
 \hline
 \end{tabular}}
 \end{table*}

 \begin{table}
 \caption{Magnetic dipole (M1) transitions widths. (LP = Linear potential model, SP = Screened potential model, NR = Non-relativistic and  RE = Relativistic, Here $E_{\gamma}$ in MeV and $\varGamma$ in KeV) \label{tab:M1CC} }
 \scalebox{0.89}{
 \begin{tabular}{cccccccccccccc}
 \hline
 \addlinespace[2pt]
 \multicolumn{2}{c}{ Transition} & \multicolumn{2}{c}{This work} &  Expt.\cite{PDGlatest}& \multicolumn{9}{c}{Other work } \\
  \addlinespace[2pt]

 Initial&Final &$E_{\gamma}$ & $\varGamma$ & $\varGamma$& \cite{Ebert:2002}   & \cite{Parmar2010} & NR(GI)\cite{Barnes:2005}& \cite{Brambilla:2004}&\cite{Eichten:2002}& \cite{Segovia:2008}&\cite{Cao:2012}& LP(SP)\cite{Deng:2016}& RE(NR)\cite{Sultan:2014}\\
 \addlinespace[3pt]
 \hline
 \addlinespace[5pt]
 $1^{3}S_{1}$ & $1^{1}S_{0}$ & 97 & 1.647  & $1.58\pm0.37$ & 1.05 & 2.01 & 2.9 (2.4)& 1.960& 1.92& 2.0& 2.2 & 2.39 (2.44)& 2.765 (2.752)\\
 $2^{3}S_{1}$ & $2^{1}S_{0}$ & 42 & 0.135  & $0.21\pm0.15$& 0.99 & 0.20 &0.21 (0.17)& 0.140& 0.04& 0.2&0.096 &0.19 (0.19) &0.198 (0.197)\\
 $3^{3}S_{1}$ & $3^{1}S_{0}$ & 36 & 0.082  & &  & 0.012 &0.046  (0.067) & & & 0.0046& 0.044 & 0.051 (0.088)&0.023 (0.044)\\
 $2^{3}S_{1}$ & $1^{1}S_{0}$ & 595 & 69.57 & $1.24\pm0.29$ & 0.95 &  &4.6  (9.6)& 0.926 & 0.91&& 3.8&8.08 (7.80)& 3.370 (4.532)\\
 $2^{1}S_{0}$ & $1^{3}S_{1}$ & 476 & 35.72 & & 1.12 &  & 7.9 (5.6)& 0.538&&7.2& 6.9&2.64 (2.29)& 5.792 (7.962) \\
 \hline
 \addlinespace[5pt]
 $1^{3}P_{2}$ & $1^{3}P_{0}$ & 97 & 1.638  &   & & & &  &&&& &\\
 $1^{3}P_{2}$ & $1^{3}P_{1}$ & 33 & 0.189  &   & & & && &&& &\\
 $1^{3}P_{2}$ & $1^{1}P_{1}$ & 22 & 0.056  &   &&  & & &&&& &\\
 $1^{1}P_{1}$ & $1^{3}P_{0}$ & 76 & 0.782 &   & & & && &&& &\\
 \hline
 \end{tabular}}
 \end{table}

 \begin{table*}
 \caption{Leptonic decay widths ({$\psi \rightarrow\varGamma_{e^{+}e^{-}}$} in KeV  ).\label{tab:annihi2e} }
 \scalebox{0.9}{
\begin{tabular}{ccccccccccccc}
\hline
 \addlinespace[2pt]
 
State & \multicolumn{2}{c}{This work} &  Expt.\cite{PDGlatest}& \multicolumn{9}{c}{Other work } \\
  \addlinespace[2pt]
 
 & $\varGamma_{l^{+}l^{-}}$ & $\varGamma_{l^{+}l^{-}}^{cf}$ &  & \cite{DSouza:2017} & \cite{Bhaghyesh:2011,Bhaghyesh:2012} & \cite{Li:2009} & \cite{Giannuzzi:2008} & \cite{Radford:2007} & \cite{Barnes:2005}&\cite{Segovia:2008}&\cite{Cao:2012}&\cite{Bhavsar:2018} \\
  \addlinespace[2pt]
 \hline
 \addlinespace[5pt]
 $J/\psi$ & 8.335 & 3.623 & $5.55\pm0.14\pm0.02$ & 3.112 & 6.847 (2.536) & 11.8 (6.60) & 4.080 & 4.28 & 12.13& 3.93 &6.0(3.3)&5.63\\
 $\psi(2S)$ & 2.496 & 1.085 & $2.33\pm0.07$ & 2.197 & 3.666 (1.358) & 4.29 (2.40) & 2.375 & 2.25 & 5.03&1.78 &2.2(1.2)&2.19\\
 $\psi(3S)$ & 1.722 & 0.748 & $0.86\pm0.07$ & 1.701 & 2.597 (0.962) & 2.53 (1.42) & 0.835 & 1.66 & 3.48&1.11 &1.8(0.98)&1.20\\
 $\psi(4S)$ & 1.378 & 0.599 & $0.58\pm0.07$ &  & 2.101 (0.778) & 1.73 (0.97) &  & 1.33 & 2.63& 0.78 &1.3(0.70)&0.63\\
 $\psi(5S)$ & 1.168 & 0.508 &  &  & 1.701 (0.633) & 1.25 (0.70) &  &  & & 0.57 & &0.24\\
 $\psi(6S)$ & 1.017 & 0.442 &  &  &  & 0.88 (0.49) &  &  & & 0.42 && \\
   \addlinespace[2pt]
 \hline 
    \addlinespace[5pt]
 $1^{3}D_{1}$ & 0.261 & 0.113 & $0.262\pm0.018$ & 0.275 & 0.096 & 0.055 (0.031) &  & 0.09 & 0.056& 0.22 &0.079(0.044)&\\
 $2^{3}D_{1}$ & 0.381 & 0.166 & $0.48\pm0.22$ & 0.223 & 0.112 & 0.066 (0.037) &  & 0.16 & 0.096& 0.30 &0.13(0.073)&\\
 $3^{3}D_{1}$ & 0.485 & 0.211 &  &  &  & 0.079 (0.044) &  &  && 0.33 && \\
 \hline
  \end{tabular}}
 \end{table*}

 \begin{table*}
  \caption{Two-photon decay widths without and with correction factor (in KeV).\label{tab:annihi2p}}
  \scalebox{0.90}{
  \begin{tabular}{cccccccccccccccc}
  \hline
  \addlinespace[2pt]
  State & \multicolumn{2}{c}{This work} &  Expt.\cite{PDGlatest}& \multicolumn{12}{c}{Other work } \\
    \addlinespace[2pt]
  
  & $\varGamma_{\gamma\gamma}$ & $\varGamma_{\gamma\gamma}^{cf}$ &  & \cite{DSouza:2017} & \cite{Negash:2015} & \cite{Bhaghyesh:2011,Bhaghyesh:2012} & \cite{Li:2009} & \cite{Laverty:2009} & \cite{Munz:1996} & \cite{Lakhina2006} & \cite{Kim:2004} & \cite{Giannuzzi:2008}&\cite{Cao:2012}&\cite{Ebert:2003b}&\cite{Munz:1996}\\
   \addlinespace[2pt]
  \hline
  \addlinespace[5pt]
 $\eta_{c}(1S)$ & 10.351 & 6.621 & $5.1\pm 0.4$ & 6.96 & 7.918 & 6.68 & 8.5 & 5.09 & 3.5 & 7.18 & 7.14 & 4.252& 7.5 & 5.5& 3.5\\
   $\eta_{c}(2S)$ & 4.501 & 2.879 & $2.15\pm0.6$ & 10.45 & 5.789 & 5.08 & 2.4 & 2.63 & 1.38 & 1.71 & 4.44 & 3.306&2.9 &1.8 &1.38\\
   $\eta_{c}(3S)$ & 3.821 & 2.444 &  & 1.03 & 0.299 & 4.53 & 0.88 &  & 0.94 & 1.21 &  & 1.992& 2.5 & &\\
   $\eta_{c}(4S)$ & 3.582 & 2.291 &  &  &  &  &  &  & 0.73 &  &  && 1.8 & &\\
   $\eta_{c}(5S)$ & 3.460  & 2.213 &  &  &  &  &  &  & 0.62 &  &  & & & &\\
   $\eta_{c}(6S)$ & 3.378 &  2.161 &  &  &  &  &  &  &  &  &  &&& & \\
     \addlinespace[2pt]
     \hline 
     \addlinespace[5pt]
   $1^{3}P_{0}$ & 1.973 & 2.015 & $2.36\pm0.35$ & 13.43 &  & 2.62 & 2.5 & 2.02 & 1.39 & 3.28 &  && 10.8 & 2.9&1.39\\
   $2^{3}P_{0}$ & 2.299 & 2.349 &  & 2.67 &  &  & 1.7 &  & 1.11 &  &  & & 6.7&1.9 &1.11\\
   $3^{3}P_{0}$ & 2.714 &  2.773 &  &  &  &  & 1.2 &  & 0.91 &  &  & & 6.5& &\\
     \addlinespace[2pt]
     \hline 
     \addlinespace[5pt]
   $1^{3}P_{2}$ & 0.526 & 0.229  & $0.53\pm0.03$ & 1.72 &  & 0.25 & 0.31 & 0.46 & 0.44 &  &  & & 0.27& 0.50&0.44\\
   $2^{3}P_{2}$ & 0.613 & 0.267 &  & 0.343 &  &  & 0.23 &  & 0.48 &  &  &&0.39&0.52 &0.48\\
   $3^{3}P_{2}$ &  0.724 &  0.315 &  &  &  &  & 0.17 &  & 0.014 &  &  && 0.66& & \\
   \hline
   \end{tabular}}
   \end{table*}

 \begin{table}
 \caption{Three-photon decay widths (in eV).\label{tab:annihi3p} }
 \centering
  \begin{tabular}{cccc}
   \hline
 \addlinespace[2pt]
  State & \multicolumn{2}{c}{This work} &  Expt.\cite{PDGlatest}  \\
  & $\varGamma_{\gamma\gamma\gamma}$ & $\varGamma_{\gamma\gamma\gamma}^{cf}$& \\
 \addlinespace[2pt]
 \hline
 \addlinespace[5pt]
 $J/\psi$ & 4.41691 & 3.94748& $1.08\pm0.032$\\
  $\psi(2S)$ & 1.83911  & 1.64365& \\
  $\psi(3S)$ & 1.55252  & 1.38752& \\
  $\psi(4S)$ & 1.45187 & 1.29756&\\
  $\psi(5S)$ & 1.40027 & 1.25145&\\
  $\psi(6S)$ &  1.36564 & 1.2205&\\
  \hline
 \end{tabular}
 \end{table}

  \begin{table}
  \caption{Three-gluon decay widths (KeV) \label{tab:annihi3g}. }
  \centering
   \begin{tabular}{cccccc}
   \hline
  \addlinespace[2pt]
   State & \multicolumn{2}{c}{This work} &  Expt.\cite{PDGlatest}& \multicolumn{2}{c}{Other work } \\
  & $\varGamma_{ggg}$  & $\varGamma_{ggg}^{cf}$ &  &\cite{Eichten:2002}& \cite{Barnes:2003}MeV\\
  \addlinespace[2pt]
  \hline
  \addlinespace[5pt]
  $J/\psi$ & 442.669  & 269.059&$59.55\pm0.18$ &$52.8\pm5$&\\
   $\psi(2S)$ & 184.318  & 112.031&$31.38\pm0.85$& $23\pm2.6$&\\
   $\psi(3S)$ & 155.596 & 94.5727&&&\\
   $\psi(4S)$ & 145.508 & 88.4413&&&\\
   $\psi(5S)$ & 140.337 & 85.2984&& &\\
   $\psi(6S)$ & 136.866 & 83.1888&&&\\
  \addlinespace[5pt]
   $1^{1}P_{1}$ & 285.127 & &&$720\pm320$&\\
   $2^{1}P_{1}$ & 420.078  && &&1.29\\
   $3^{1}P_{1}$ & 558.78 &&&& \\
  \addlinespace[5pt]
   $1^{3}D_{1}$ & 189.367 &&&216 &1.15\\
   $2^{3}D_{1}$ & 359.346 &&&& \\
   $3^{3}D_{1}$ &  556.588 &&&&\\
  \addlinespace[5pt]
   $1^{3}D_{2}$ & 53.8761 &&& 36&0.08\\
   $2^{3}D_{2}$ & 102.236  && &&\\
   $3^{3}D_{2}$ & 158.353 && &&\\
    \addlinespace[5pt]
   $1^{3}D_{3}$ & 89.7001 &&& 102&0.18\\
   $2^{3}D_{3}$ & 170.217  && &&\\
   $3^{3}D_{3}$ & 263.647 & &&&\\
   \hline
  \end{tabular}
   \end{table}

 \begin{table*}
 \caption{Two-gluon decay widths(in MeV).\label{tab:annihi2g}}
 \centering
 \begin{tabular}{cccccccccc}
 \hline
 \addlinespace[2pt]
  State & \multicolumn{2}{c}{This work} &  Expt.\cite{PDGlatest}& \multicolumn{6}{c}{Other work } \\
  & $\varGamma_{gg}$& $\varGamma_{gg}^{cf}$ &  & \cite{DSouza:2017} & \cite{Negash:2015} & \cite{Bhaghyesh:2011,Bhaghyesh:2012}& \cite{Laverty:2009} & \cite{Kim:2004}& \cite{Eichten:2002} \\
 \addlinespace[2pt]
 \hline
 \addlinespace[5pt]
 $\eta_{c}(1S)$ & 24.249 & 36.587 & $28.6\pm2.2$ & 28.60 & 13.070 & 32.44 & 15.70 & 19.6& $17.4\pm2.8$\\
  $\eta_{c}(2S)$ & 10.545  & 15.910 & $14\pm7$ & 42.90 & 9.534 & 24.64 & 8.10 & 12.1& $8.3\pm1.3$\\
  $\eta_{c}(3S)$ & 8.952 & 13.507 &  & 4.26 & 4.412 & 21.99 &  & & \\
  $\eta_{c}(4S)$ & 8.392 & 12.662 &  &  &  &  &  & & \\
  $\eta_{c}(5S)$ & 8.106 & 12.230 &  &  &  &  &  && \\
  $\eta_{c}(6S)$ & 7.914 &  11.941 &  &  &  &  &  && \\
   \addlinespace[2pt]
   \hline 
   \addlinespace[5pt]
  $1^{3}P_{0}$ & 4.621 & 9.274 & $10\pm0.6$ & 47.76 &  & 15.67 & 4.68 & &$14.3\pm3.6$ \\
  $2^{3}P_{0}$ & 5.386  & 10.810 &  & 9.50 &  &  &  & &\\
  $3^{3}P_{0}$ & 6.357 & 12.758 &  &  &  &  &  && \\
    \addlinespace[2pt]
    \hline 
    \addlinespace[5pt]
  $1^{3}P_{2}$ & 1.232  & 0.945  & $1.97\pm0.11$ & 5.27 &  & 1.46 & 1.72 && $1.71\pm0.21$ \\
  $2^{3}P_{2}$ & 1.436  & 1.101 &  & 1.04 &  &  &  && \\
  $3^{3}P_{2}$ & 1.695 & 1.300 &  &  &  &  &  && \\
   \addlinespace[2pt]
   \hline 
   \addlinespace[5pt]
  $1^{1}D_{2}$ & 12.460 (KeV) &  &  &  &  &  &  & & 110 (KeV)\\
  $2^{1}D_{2}$ & 21.679 (KeV) &  &  &  &  &  &  && \\
  $3^{1}D_{2}$ & 31.757 (KeV) &  &  &  &  &  &  && \\
  \hline
 \end{tabular}
  \end{table*}

  \begin{table}
  \caption{$n^{3}S_{1}\rightarrow\gamma gg$ decay widths.\label{tab:annihip2g}}
 \centering
  \begin{tabular}{cccc}
  \hline
 \addlinespace[2pt]
  State & \multicolumn{2}{c}{This work} &  Expt.\cite{PDGlatest} \\
  State & $\varGamma_{\rightarrow\gamma gg}$ (KeV) & $\varGamma_{\rightarrow\gamma gg}^{cf}$ (KeV)&  \\
  \addlinespace[2pt]
   \hline 
   \addlinespace[3pt]
  $J/\psi$ &31.0421 &  8.99657 &$8.18\pm0.25$\\
  $\psi(2S)$&  12.9253 &  3.74599& $2.93\pm0.16$ \\
  $\psi(3S)$&  10.9111&  3.16224& \\
  $\psi(4S)$ & 10.2037&  2.95723&\\
  $\psi(5S)$&  9.8411 & 2.85214&\\
  $\psi(6S)$&   9.59771 &  2.7816&\\
  \hline
  \end{tabular}
  \end{table} 
  
  \begin{table}
  \caption{$n^{3}P_{1}\rightarrow q\overline{q}+g$ decay widths.\label{tab:annihiqqg}}
 \centering
  \begin{tabular}{cc}
  \hline
   State & This work  \\
  & $\varGamma_{q\overline{q}+g}$(KeV)\\
  \addlinespace[2pt]
   \hline 
   \addlinespace[3pt]
  $1^{3}P_{1}$ & 342.152 \\
  $2^{3}P_{1}$ & 504.093 \\
  $3^{3}P_{1}$ & 670.536\\
  \hline
  \end{tabular}
  \end{table}

 \begin{multicols}{2}
 
  \section{Results and Discussion\label{sec:Resu}}
 
{\bf In the framework of Cornell potential with a Gaussian wave function and relativistic correction of the Hamiltonian, comprise with a $\mathcal{O}(1/m)$ rectification in the potential energy term and elaboration of the kinetic energy term up to ${{\cal{O}}\left({\bf p}^{10}\right)}$,} we have studied the mass spectra of charmonium states. {\bf We have calculated center of weight masses (value of Hamiltonian yields) for the nS $(n\leq6)$, $ nP\; \text{and}\; nD \;(n\leq3)$ charmonium states and tabulated in Table(\ref{tab:ccmsa}).}
We observed that Hamiltonian yields for nS $(n\leq3)$ and  $ nP\; \text{and}\; nD \;(n\leq3)$ are in accordance with experimental as well as values predicted by other theoretical model, whereas for nS $(4\leq n\leq6)$ are underestimated and/or overestimated compared to results of other theoretical model. 

The calculated mass of charmonium states are {\bf graphically} represented in  Fig.\ref{fig:MassCC}  and tabulated in Table(\ref{tab:massescc}) with experimentally observed results. After addition of the spin hyperfine interaction in fixed spin average mass for {\bf the ground} state, we obtained pseudoscalar state mass $\eta_{c}$ (2995 MeV) and vector state mass $J/\psi$ (3094 MeV). 
The estimated mass $2^{1}S_{0}$ (3606 MeV) is 33 MeV lower than experimentally observed mass, whereas mass $3^{3}S_{1}$(4036) is accordance with mass given in PDG \cite{PDGlatest} and other model estimates \cite{Deng:2016,Ebert2010,Li:2009}. Our calculated mass $5^{3}S_{1}$ (4654 MeV) is 11 MeV higher than value quoted in PDG \cite{PDGlatest} and accordance with mass estimated by other model \cite{Radford:2007,Sultan:2014}. We have assigned $X(4660)$ to the $5^{3}S_{1}$ state of charmonium. Estimated mass of $6^{3}S_{0}$ (4893 MeV)  and $6^{3}S_{1}$ (4925 MeV) states are agreement with mass estimated by other model \cite{Sultan:2014}.

The P-wave states,  $1^{3}P_{1}$ with predicted mass 3511 MeV,  $1^{1}P_{1}$ with predicted mass 3525 MeV and $2^{3}P_{2}$ with predicted mass 3556 MeV are in good agreement with experimental observed value \cite{PDGlatest}. 

{\bf We have assigned newly observed charmonium like state $X(3900)$ to  the $2^{1}P_{1}$ (3936 MeV) and state $X(3872)$ to the $2^{3}P_{1}$ (3925 MeV). The mass predicted for state $2^{1}P_{1}$ (3936 MeV) and state $2^{3}P_{1}$ (3925 MeV) is in good agreement with mass predicted by other model \cite{Deng:2016,Yang:2015cc,Ebert2010,Barnes:2005,Cao:2012,Ebert2002,Sultan:2014}. The candidate $X(3872)$ as the $2^{3}P_{1}$ state with well established quantum numbers, although the interpretation of it as a
molecular state \cite{Tornqvist:2004,Braaten:2007} and was questioned in Ref.\cite{Albaladejo:2017}, while Ref.\cite{Hanhart:2007} interpreted it as virtual state}.

We have also assigned charmonium like states, $X(3915)$ and $X(4274)$ to the $2^{3}P_{0}$ (3866 MeV) and $3^{3}P_{1}$(4257 MeV) states respectively. {\bf To consider $X(3915)$ as the  $2^{3}P_0$ state is still problematic and was also pointed out in Ref.\cite{Deng:2016,Zhao:2013} and the references therein. In Ref.\cite{Zhao:2013,Liu:2009fe,Guo:2010}, the authors suggest the $X(3915)$ as the  $2^{3}P_0$ state faces the following problems: First, A scalar meson should be the open-flavor modes for the  dominant decay channels, above the corresponding thresholds. The Facts that  $X(3915)$ can couple in an S-wave and the $D\bar{D}$ channel, although was not observed in the $D\bar{D}$ channel. Second, the mass splitting between the state $1^{3}P_2$ and  $1^{3}P_0$ is 141 MeV, while the mass splitting between relatively well determined $X(3930)$ as the $2^{3}P_2$ state and  $X(3915)$ as the $2^{3}P_0$ state is 9 MeV, which is too small for the hyperfine splitting}. 

We observed that new charmonium like states $X(4140)$ and $X(4274)$ with their quantum number $J^{PC}=1^{++}$ is a good candidate for $3^{3}P_{1}$ {\bf state} within screen potential model and linear potential model respectively. However none of the {\bf models} can give $J^{PC}=1^{++}$  charmonium state masses 4147 MeV and 4273 MeV at the same time, which may indicate the exotic nature of $X(4140)$ and/or $X(4274)$, which was also pointed out in Ref.\cite{Deng:2016}.
  
The predicted mass for $1^{3}D_{1}$ (3799 MeV), $1^{3}D_{2}$ (3805 MeV) and  $2^{3}D_{1}$ (4145 MeV) states are accordance with Experiment observed results \cite{PDGlatest} as well as with good agreement with other model prediction. \cite{Deng:2016,Ebert2010,Barnes:2005,Cao:2012,Li:2009,Ebert2002,Sultan:2014}. The estimated masses of charmonium using our model are overall in agreement (with few MeV difference) with experimentally observed values. It is found that states with a mass of $M<4.1$ GeV are in good agreement with other theoretical estimates. 
  
Table(\ref{tab:masssplit}) shows the hyperfine splittings for S wave states and fine splittings for some P wave states. For comparison, the experimental data from the PDG \cite{PDGlatest} and predictions with other theoretical model are listed in the same table as well. we observed that the predicted hyperfine splittings, up-to 2S states are in agreement with the world average data \cite{PDGlatest} and predictions with other theoretical model. The hyperfine splittings for 3S to 6S states have a different value in the different theoretical model. By comparing our predicted results with other theoretical model, we observed that masses of the low-lying  nS $(n\leq2)$, nP, nD $(n=1)$, charmonium states are in less difference, whereas masses of  the higher charmonium states  $nS\;(n\geq3)$, $nP,\;nD \;(n\geq2)$, are in notable difference.
  
The estimated pseudoscalar and vector decay constant $f_{P}$($f_{Pcor}$) and $f_{V}$($f_{Vcor}$) respectively, without(with) QCD correction are tabulated in Table(\ref{tab:decaycc}), which are in  agreement  with experimental results as well as other theoretical model estimates.
  
We calculate radiative E1 and M1 dipole transitions widths and are tabulated in Tables(\ref{tab:E1CC}, \ref{tab:M1CC}). We calculate the E1 transition of  $\varGamma[1P \rightarrow (1S)\gamma]$, $\varGamma[2S \rightarrow (1P)\gamma]$, $\varGamma[1D \rightarrow (1P)\gamma]$, $\varGamma[2P \rightarrow (2S)\gamma]$ and $\varGamma[2P \rightarrow (1D)\gamma]$ using the masses predicted by our model. Our calculated E1 transition of  $\varGamma[1P \rightarrow (1S)\gamma]$ and $\varGamma[2S \rightarrow (1P)\gamma]$ are lesser than experimental results as well as other theoretical estimates, whereas for $\varGamma[1D \rightarrow (1P)\gamma]$, $\varGamma[2P \rightarrow (2S)\gamma]$ and $\varGamma[2P \rightarrow (1D)\gamma]$ transition, are in agreement with the estimates of other theoretical model. Our prediction of $\varGamma[1^{3}D_1 \rightarrow (1^{3}P_1)\gamma]$ and $\varGamma[1^{3}D_1 \rightarrow (1^{3}P_0)\gamma]$ almost double in comparison with the PDG average data \cite{PDGlatest} while  prediction of $\varGamma[1^{3}D_1 \rightarrow (1^{3}P_2)\gamma]$ is in agreement with the PDG average data \cite{PDGlatest} as well as with predicted by other model.
  
We also, calculate the M1 transition of the low-lying 1S, 2S and 3S states as well as 1P states. Our prediction of $\varGamma[1^{3}S_1 \rightarrow (1^{1}S_0)\gamma]$ and $\varGamma[2^{3}S_1 \rightarrow (2^{1}S_0)\gamma]$ are in agreement with the PDG average data \cite{PDGlatest}, while $\varGamma[2^{3}S_1 \rightarrow (1^{1}S_0)\gamma]$ is much larger than the PDG average data \cite{PDGlatest}. {\bf Gang Li and Qiang Zhao, Ref.\cite{Li:2007,Li:2011a} studied intermediate meson loop contributions to $1^{3}S_1, 2^{3}S_1 \rightarrow \gamma 2^{1}S_0,(\gamma1^{1}S_0)$ apart from
the dominant M1 transitions in an effective Lagrangian approach. Results shows that the IML contributions are relatively small but play a crucial role. Radiative decay widths including the M1 in the GI model and intermediate hadronic loops for $1^{3}S_1\rightarrow \gamma 2^{1}S_0$ is $1.59$ KeV and for $2^{3}S_1 \rightarrow \gamma 2^{1}S_0(\gamma1^{1}S_0)$ is 0.032(0.86) KeV \cite{Li:2007}, whereas results including the M1 transition amplitude of the GI model and IML transitions for $1^{3}S_1\rightarrow \gamma 2^{1}S_0$ is $1.58\pm0.37$ KeV and for $2^{3}S_1 \rightarrow \gamma 2^{1}S_0(\gamma1^{1}S_0)$ is $0.08\pm0.03$ ($2.78^{+2.65}_{-1.75}$) KeV \cite{Li:2011a}.}

Our prediction of $\varGamma[3^{3}S_1 \rightarrow (3^{1}S_0)\gamma]$ is in agreement with the other theoretical model prediction, while prediction of $\varGamma[2^{1}S_0 \rightarrow (1^{3}S_1)\gamma]$ is larger than the predicted by other theoretical model. We observed that the various models have different estimates of E1 and M1 transitions, it may be due to the different models have different parameters or  treatments in the relativistic corrections. The E1 and  M1 transitions as a whole are strongly model dependence and more studies are required in both experiments as well theory.
    
{\bf We estimate partial decay width $\varGamma$ and $\varGamma^{cf}$ (with QCD correction factor) of annihilation processes,  using the masses predicted by our potential model and the radial wave function at the origin, for $e^{+}e^{-}$, two-photon, three-photon, two-gluon, three-gluon, $\gamma gg$ and $q\bar{q}+g$ are tabulated in Tables(\ref{tab:annihi2e}-\ref{tab:annihiqqg}) and are {\bf compared} with experimental results from PDG\cite{PDGlatest} as well as other theoretically calculated estimates.}
  
We observed that our estimated leptonic decay without QCD correction for $J/\psi$, $\psi (2S)$, $\psi (3S)$ and $\psi (4S)$ is higher than experimentally observed leptonic decay width. {\bf After QCD correction, estimated leptonic decay} is 1.93 KeV, 1.24 KeV, 0.11 KeV and 0.019 KeV lesser than the experimental result  for $J/\psi$, $\psi (2S)$, $\psi (3S)$ and $\psi (4S)$ state respectively. Also, our estimated leptonic decay with QCD correction for $n^{3}D_1$ state is much lower than the experimental result.

{\bf Our estimated two-photon and two-gluon decay widths with QCD correction for $\eta_c (nS)$, $n^{3}P_0$ and $n^{3}P_2$ state are  accordance with experimentally observed results as well as with the other theoretical estimates. Our estimated three-photon decay widths with QCD correction for $J/\psi$ is lower than the experimentally observed result whereas estimated three-gluon decay widths with QCD correction for $J/\psi$ and $\psi(2S)$ state is higher than the experimentally observed result as well as other theoretical estimates.}

Our estimated $\gamma gg$ decay width with QCD correction for $J/\psi$ and $\psi(2S)$ state is {\bf accordance} with the experimentally observed result. We have also {\bf compute} $q\bar{q}+g$ decay width for $n^{3}P_1$ states.  We observed that radiative QCD corrections  modify theoretical predictions considerably and  {\bf bring} estimated result close to experimental data. We also observed that the estimated values of annihilation decay width by of various models show a wide range of variations.  Due to the considerable uncertainties arise from the wave functions dependence model and possible relativistic {\bf as well as} QCD radiative corrections, we would like to mention that formulas used for  calculation of annihilation decay width should be regarded as estimates of the partial widths rather than precise predictions.

 \subsection{Regge trajectories \label{sec:reg}}
   
We plot the Regge trajectories for the $(n,M^{2})$ and  $(J,M^{2})$ planes with the help of masses estimated by our potential model. The "daughter" trajectories are the trajectories with the same value of $J$ and differ by a quantum number correspondent to the radial quantum number. The masses of the "daughter" trajectories are higher than those for the leading trajectory with given quantum numbers. The linearity of Regge trajectories {\bf represents} as a  reflection of strong forces between quarks at large distances (color confinement).

\begin{figure}
        \centering
        \includegraphics[bb=30bp 60bp 750bp 550bp,clip,width=0.80\textwidth]{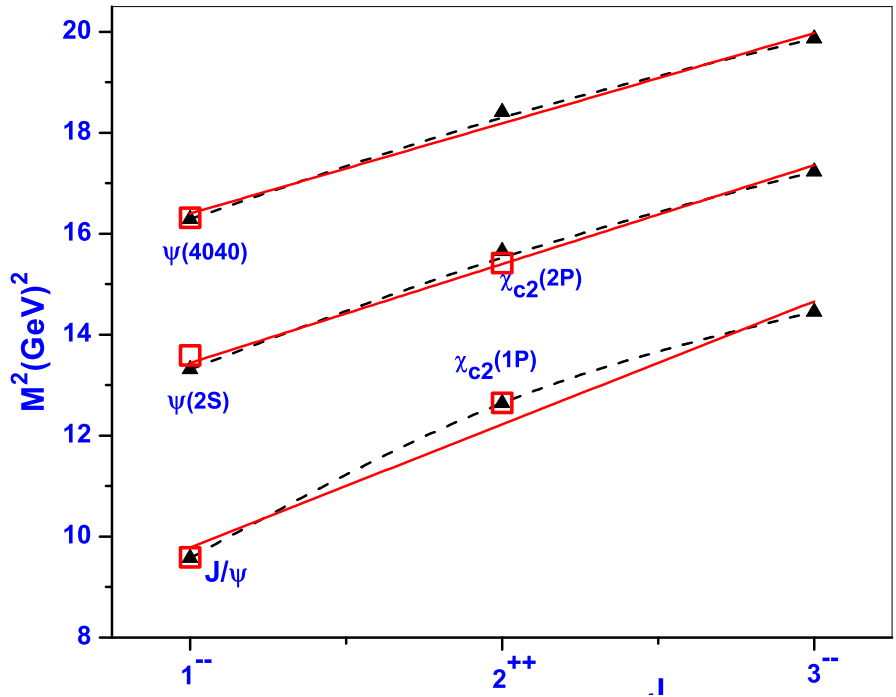}
        \caption{Regge trajectory ($M^{2}\rightarrow J$) with natural parity. \label{fig:NPmesonCC}}
   \end{figure}

\begin{figure}
      \centering
      \includegraphics[bb=30bp 60bp 750bp 550bp,clip,width=0.80\textwidth]{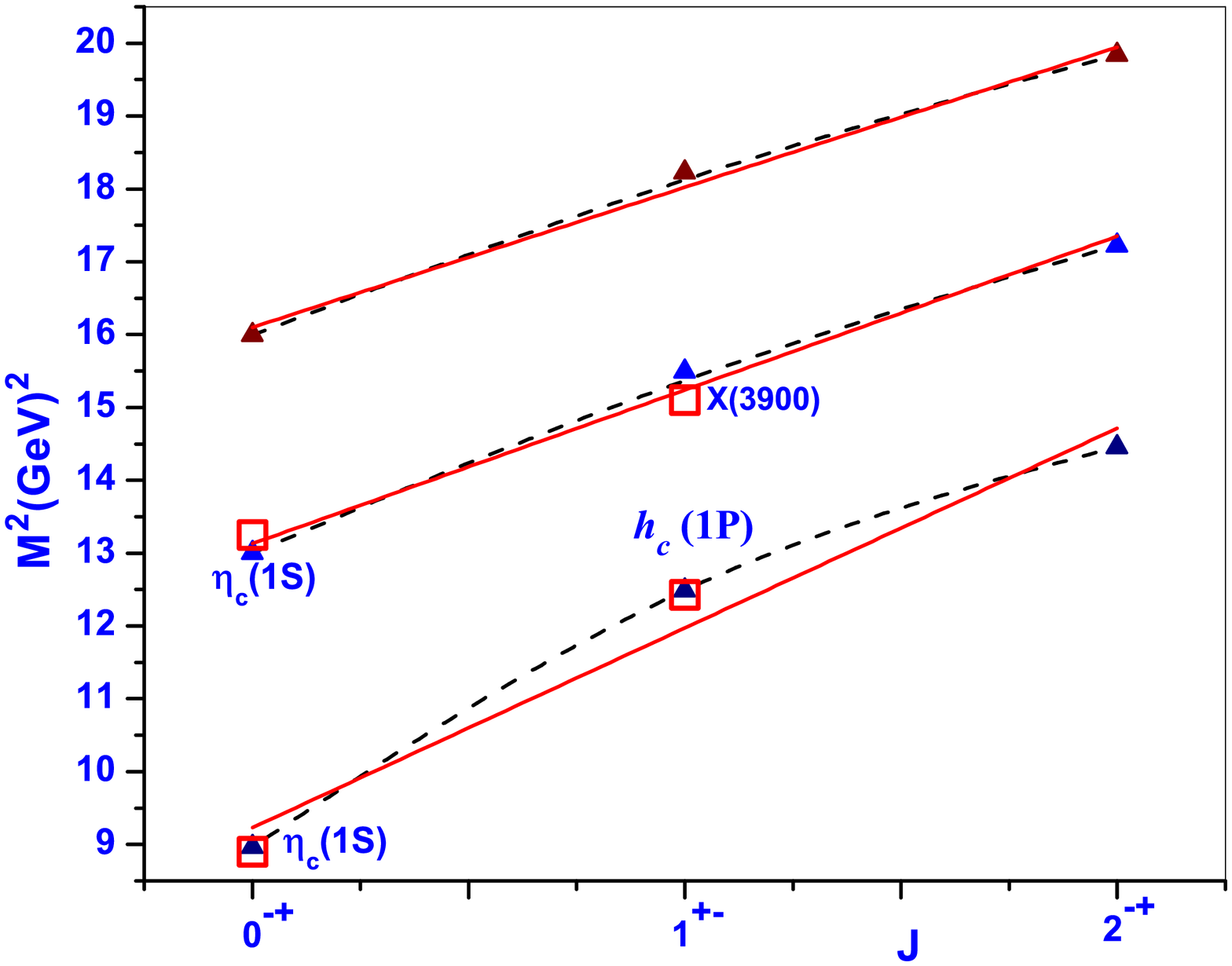}
    \caption{Regge trajectory ($M^{2}\rightarrow J$) with unnatural parity. \label{fig:UNPmesonCC}}
     \end{figure}

\begin{figure}
     \centering
      \includegraphics[bb=30bp 60bp 750bp 550bp,clip,width=0.80\textwidth]{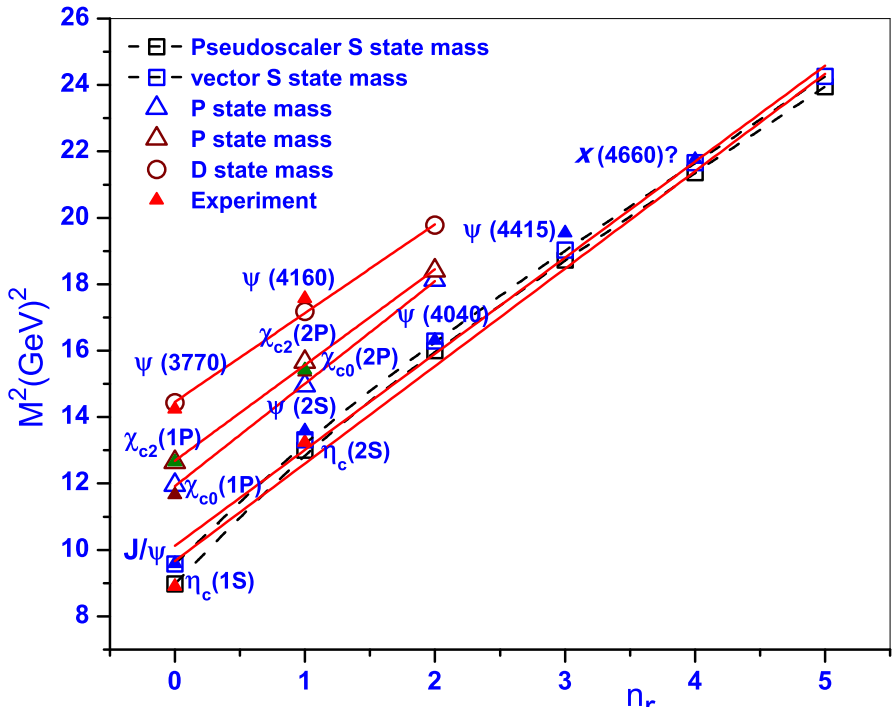}
      \caption{Regge trajectory ($M^{2}\rightarrow n_{r}$) for the pseudoscalar and vector $S$ state and excited $P$ and $D$ state masses.\label{fig:PsVmesonCC}}
     \end{figure}

\begin{figure}
      \centering
      \includegraphics[bb=30bp 60bp 750bp 550bp,clip,width=0.80\textwidth]{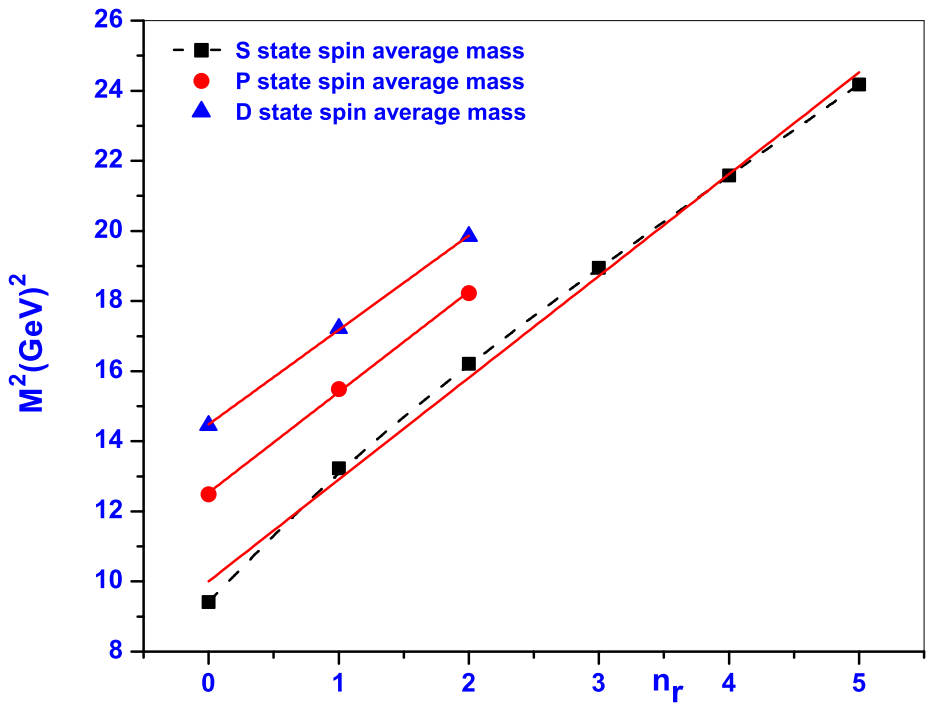}
      \caption{Regge trajectory ($M^{2}\rightarrow n_{r}$) for the S-P-D states center of weight mass.\label{fig:SavmesonCC}}
      \end{figure}

The Regge trajectories in the $(J,M^{2})$ plane with  $(P=(-1)^{J})$ ($J^{P}=1^{-},2^{+},3^{-}$ ) natural and $(P=(-1)^{J-1})$ ($J^{P}=0^{-},1^{+},2^{-}$ ) unnatural parity are depicted in Figs.~(\ref{fig:NPmesonCC}-\ref{fig:UNPmesonCC}). {\bf In figure, charmonium masses estimated by our model are represented by the solid triangles whereas experimentally available mass with the corresponding charmonium name are represented by hollow squares.} The Regge trajectories for $n_{r}= n-1$  principal quantum number in the $(n_{r},M^{2})$ plane are describe in Figure~(\ref{fig:PsVmesonCC}) and  Figure~(\ref{fig:SavmesonCC}).
  
The following definitions are used to calculate the $\chi^{2}$  fitted slopes ($\alpha$, $\beta$) and the intercepts ($\alpha_{0}$, $\beta_{0}$) \cite{Kher:2017,Kher:2017b}.  
  \begin{equation}
        J=\alpha M^{2}+\alpha_{0}.\label{eq:J regge}
  \end{equation} 
 \begin{equation}
        n_{r}=\beta M^{2}+\beta_{0}\label{eq:nr regge}
  \end{equation}

Calculated slopes and intercepts are tabulated in Tables~(\ref{tab:alfaCC},\ref{tab:bitaCC},\ref{tab:SpinaveCC}). The estimated masses of  the charmonium fit well to  the $(n,M^{2})$ and  $(J,M^{2})$ planes  trajectories. The daughter trajectories, which involve both  radially and orbitally excited states, turn out to be almost linear, equidistant and parallel whereas The parent Regge trajectories, which start from ground  states, are exhibiting a nonlinear behavior in the lower mass region in both planes.
  
We observed that the linearity of the Regge trajectories depends on quark masses, as the  orbital momentum $\ell$ of the state is proportional to its mass: $\ell=\alpha M^{2}(\ell)+\alpha(0)$, where the slope $\alpha$ depends on the flavor content of the states lying on the corresponding trajectory. In the Regge phenomenology, the radial spectrum of heavy quarkonia typically {\bf leads} to strong nonlinearities, in the framework of hadron string model \cite{Afonin:2016}.

\end{multicols}

   \begin{table}
  \caption{Slopes and intercepts of the $(J,\: M^{2})$ Regge trajectories with unnatural and natural parity.\label{tab:alfaCC} }
  \centering
  \begin{tabular}{cccc}
  \hline
 \addlinespace[2pt]    
 {Parity}  & {Trajectory} & {$\alpha(GeV^{-2})$} & {$\alpha_{0}$}\\
  \addlinespace[2pt]    
       \hline
  \addlinespace[3pt]
  \multirow{3}{*}{Unnatural } & Parent & $0.355\pm0.058$ & $-3.252\pm0.706$\\
  & First daughter & $0.471\pm0.038$ & $-6.164\pm0.576$\\
  & Second daughter & $0.518\pm0.032$ & $-8.319\pm0.570$\\
        \addlinespace[2pt]    
      \hline
       \addlinespace[3pt]
 \multirow{3}{*}{Natural}  & Parent &  $0.401\pm0.060$ & $-2.902\pm0.746$\\
 & First daughter &  $0.504\pm0.057$ & $-5.764\pm0.877$\\
 & Second daughter & $0.553\pm0.059$ & $-8.057\pm1.081$\\
 \hline
 \end{tabular}
 \end{table}

    \begin{table}
   \caption{Slopes and intercepts for the $(n_{r},\: M^{2})$ Regge trajectories. \label{tab:bitaCC} }
  \centering 
  \begin{tabular}{cccc}
  \hline
  \addlinespace[2pt]    
   Meson & $J^P$ & $\beta(GeV^{-2})$ & $\beta_{0}$\\
   \addlinespace[2pt]    
        \hline
   \addlinespace[3pt]
 $\eta_{c}$ & $0^{-+}$ & $0.341\pm0.017$ & $-3.236\pm0.303$\\
 $\Upsilon$ & $1^{--}$ & $0.347\pm0.014$ & $-3.463\pm0.252$\\
 $\chi_{c0}$ & $0^{++}$ & $0.324\pm0.006$ & $-3.861\pm0.088$\\
 $\chi_{c1}$ & $1^{++}$ & $0.355\pm0.007$ & $-4.441\pm0.112$\\
 $h_{c}$ & $1^{+-}$ & $0.346\pm0.009$ & $-4.399\pm0.138$\\
 $\chi_{c2}$ & $2^{++}$ & $0.345\pm0.012$ & $-4.284\pm0.183$\\
 $\psi(^{3}D_{1})$ & $1^{--}$ & $0.374\pm0.006$ & $-5.406\pm0.104$\\
 $\psi(^{3}D_{2})$ & $2^{--}$ & $0.377\pm0.009$ & $-5.473\pm0.159$\\
  $\psi(^{1}D_{2})$ & $2^{-+}$ & $0.371\pm0.006$ & $-5.372\pm0.101$\\
 $\psi(^{3}D_{3})$ & $3^{--}$ & $0.369\pm0.006$ & $-5.344\pm0.100$\\
 \hline
  \end{tabular}
   \end{table}

   \begin{table}
   \caption{Slopes and intercepts of  $(n_{r},\: M^{2})$ Regge trajectory for center of weight mass.\label{tab:SpinaveCC} }
  \centering
   \begin{tabular}{ccc}
   \hline
         \addlinespace[2pt]  
   Trajectory & $\beta(GeV^{-2})$ & $\beta_{0}$ \\
   \addlinespace[2pt]    
   \hline
   \addlinespace[5pt]
  S State & $0.342\pm0.012$ & $-3.413\pm0.226$ \\
  P State & $0.348\pm0.009$ & $-4.36\pm0.1464$ \\
  D State & $0.371\pm0.006$ & $-5.372\pm0.101$ \\
  \hline
  \end{tabular}
  \end{table}

 \begin{multicols}{2}

\section{Conclusion\label{sec:conclusion}}

We can conclude from the mass spectra of charmonium, Tables~(\ref{tab:ccmsa},\ref{tab:massescc}),  investigated using a Cornell potential with relativistic correction to the Hamiltonian, are accordance with the available experimental results as well as predicted by the other theoretical model. The predicted pseudoscalar $(f_{Pcor})$ and the vector $(f_{Vcor})$ decay constants with QCD correction using our estimated charmonium masses are in accordance with experimental as well as predicted by other theoretical model.

We observed from the Regge trajectories Figs.~(\ref{fig:NPmesonCC}-\ref{fig:SavmesonCC}), that the experimental masses of charmonium  states are  sitting nicely.  In the mass region of the lowest excitations of charmonium, the slope of the trajectories decreases with increasing quark mass. The curvature of the trajectory near the ground state is due to the contribution of the color Coulomb interaction, which increases with mass. Hence, the Regge trajectories of the charmonium are basically nonlinear and exhibiting a nonlinear behavior in the lower mass region.  

From a comparison of our estimated radiative (E1 and M1 dipole) transitions width with other theoretical estimations, we conclude that the various models have very different predictions of E1 and M1 dipole transitions may be due to different parameters and treatments are used in the relativistic corrections in the model. The calculated E1 and M1 dipole transitions width using the masses and parameters estimated by our model are in agreement with other theoretical and experimental predictions. Although, in most cases, more precise experimental measurements are required.

We also conclude from calculated annihilation decay widths using the Van Royen-Weisskopf relation, that the inclusion of QCD correction factors is helpful to {\bf bring} estimated results close to experimental results. The various models show a wide range of variations in results of annihilation decay widths, which may be resolved using the NRQCD (non-relativistic QCD) and pNRQCD (potential
non-relativistic QCD)formalism. \\

{\bf Acknowledgements} A. K. Rai acknowledge the financial support extended by Department of Science of Technology, India  under SERB fast track scheme SR/FTP /PS-152/2012.\\

\bibliographystyle{epj}
\bibliography{myrefcc}

\begin{thebibliography}{133}

\bibitem{Aubert:1974}
J.J. Aubert et~al. (E598), Phys. Rev. Lett. \textbf{33}, 1404 (1974)

\bibitem{Augustin:1974}
J.E. Augustin et~al. (SLAC-SP-017), Phys. Rev. Lett. \textbf{33}, 1406 (1974)

\bibitem{Brambilla2011}
N.~Brambilla, S.~Eidelman, B.~Heltsley, R.~Vogt, G.~Bodwin et~al., Eur. Phys.
  J. \textbf{C71}, 1534 (2011)

\bibitem{PDGlatest}
C.~Patrignani, P.D. Group, Chinese Physics C \textbf{40}, 100001 (2016)

\bibitem{Siegrist:1976}
J.~Siegrist et~al., Phys. Rev. Lett. \textbf{36}, 700 (1976)

\bibitem{Brandelik:1978}
R.~Brandelik et~al. (DASP), Phys. Lett. \textbf{76B}, 361 (1978)

\bibitem{Eichten1980}
E.~Eichten, K.~Gottfried, T.~Kinoshita, K.D. Lane, T.M. Yan, Phys. Rev. D
  \textbf{21}, 203 (1980)

\bibitem{Aaij2013}
R.~Aaij et~al. (LHCb), Phys. Rev. Lett. \textbf{111}, 101805 (2013)

\bibitem{Wang:2007}
X.L. Wang et~al. (Belle), Phys. Rev. Lett. \textbf{99}, 142002 (2007)

\bibitem{Pakhlova:2008}
G.~Pakhlova et~al. (Belle), Phys. Rev. Lett. \textbf{101}, 172001 (2008)

\bibitem{Rapidis:1977}
P.A. Rapidis et~al., Phys. Rev. Lett. \textbf{39}, 526 (1977)

\bibitem{Bacino:1977}
W.~Bacino et~al., Phys. Rev. Lett. \textbf{40}, 671 (1978)

\bibitem{Abrams:1979}
G.S. Abrams et~al., Phys. Rev. \textbf{D21}, 2716 (1980)

\bibitem{Ablikim:2006}
M.~Ablikim et~al. (BES), Phys. Lett. \textbf{B652}, 238 (2007)

\bibitem{Anashin:2011}
V.V. Anashin et~al., Phys. Lett. \textbf{B711}, 292 (2012)

\bibitem{Abe:2007a}
K.~Abe et~al. (Belle), Phys. Rev. Lett. \textbf{98}, 082001 (2007)

\bibitem{Abe:2007b}
P.~Pakhlov et~al. (Belle), Phys. Rev. Lett. \textbf{100}, 202001 (2008)

\bibitem{Sreethawong:2013}
W.~Sreethawong, K.~Xu, Y.~Yan (2013), \texttt{1306.2780}

\bibitem{Wang:2016}
Z.H. Wang, Y.~Zhang, L.~Jiang, T.H. Wang, Y.~Jiang, G.L. Wang, Eur. Phys. J.
  \textbf{C77}(1), 43 (2017)

\bibitem{Bhardwaj:2013}
V.~Bhardwaj et~al. (Belle), Phys. Rev. Lett. \textbf{111}(3), 032001 (2013)

\bibitem{Ablikim:2015a}
M.~Ablikim et~al. (BESIII), Phys. Rev. Lett. \textbf{115}(1), 011803 (2015)

\bibitem{Choi:2003}
S.K. Choi et~al. (Belle), Phys. Rev. Lett. \textbf{91}, 262001 (2003)

\bibitem{Acosta:2003}
D.~Acosta et~al. (CDF), Phys. Rev. Lett. \textbf{93}, 072001 (2004)

\bibitem{Abazov:2004}
V.M. Abazov et~al. (D0), Phys. Rev. Lett. \textbf{93}, 162002 (2004)

\bibitem{Aubert:2004}
B.~Aubert et~al. (BaBar), Phys. Rev. \textbf{D71}, 071103 (2005)

\bibitem{Choi:2011}
S.K. Choi et~al., Phys. Rev. \textbf{D84}, 052004 (2011)

\bibitem{Aaltonen:2009}
T.~Aaltonen et~al. (CDF), Phys. Rev. Lett. \textbf{103}, 152001 (2009)

\bibitem{Aubert:2008gu}
B.~Aubert et~al. (BaBar), Phys. Rev. \textbf{D77}, 111101 (2008)

\bibitem{Abulencia:2006}
A.~Abulencia et~al. (CDF), Phys. Rev. Lett. \textbf{98}, 132002 (2007)

\bibitem{Ablikim:2013c}
M.~Ablikim et~al. (BESIII), Phys. Rev. Lett. \textbf{112}(9), 092001 (2014)

\bibitem{Barnes:2003}
T.~Barnes, S.~Godfrey, Phys. Rev. \textbf{D69}, 054008 (2004)

\bibitem{Choi:2004}
S.K. Choi et~al. (Belle Collaboration), Phys. Rev. Lett. \textbf{94}, 182002
  (2005)

\bibitem{delAmoSanchez:2010}
P.~del Amo~Sanchez et~al. (BaBar), Phys. Rev. \textbf{D82}, 011101 (2010)

\bibitem{Lees:2012a}
J.P. Lees et~al. (BaBar), Phys. Rev. \textbf{D86}, 072002 (2012),
  \texttt{1207.2651}

\bibitem{Liu:2009}
X.~Liu, Z.G. Luo, Z.F. Sun, Phys. Rev. Lett. \textbf{104}, 122001 (2010)

\bibitem{Zhou:2015}
Z.Y. Zhou, Z.~Xiao, H.Q. Zhou, Phys. Rev. Lett. \textbf{115}(2), 022001 (2015)

\bibitem{Uehara:2005}
S.~Uehara et~al. (Belle), Phys. Rev. Lett. \textbf{96}, 082003 (2006)

\bibitem{Aubert:2010}
B.~Aubert et~al. (BaBar), Phys. Rev. \textbf{D81}, 092003 (2010)

\bibitem{Ablikim:2013b}
M.~Ablikim et~al. (BESIII), Phys. Rev. Lett. \textbf{110}, 252001 (2013)

\bibitem{Liu:2013}
Z.Q. Liu et~al. (Belle), Phys. Rev. Lett. \textbf{110}, 252002 (2013)

\bibitem{Xiao:2013}
T.~Xiao, S.~Dobbs, A.~Tomaradze, K.K. Seth, Phys. Lett. \textbf{B727}, 366
  (2013)

\bibitem{Ablikim:2015b}
M.~Ablikim et~al. (BESIII), Phys. Rev. \textbf{D92}(9), 092006 (2015)

\bibitem{Aaltonen:2009a}
T.~Aaltonen et~al. (CDF), Phys. Rev. Lett. \textbf{102}, 242002 (2009)

\bibitem{Chatrchyan:2013}
S.~Chatrchyan et~al. (CMS), Phys. Lett. \textbf{B734}, 261 (2014)

\bibitem{Abazov:2015}
V.M. Abazov et~al. (D0), Phys. Rev. Lett. \textbf{115}(23), 232001 (2015)

\bibitem{Abazov:2013}
V.M. Abazov et~al. (D0), Phys. Rev. \textbf{D89}(1), 012004 (2014)

\bibitem{Liu:2009a}
X.~Liu, S.L. Zhu, Phys. Rev. \textbf{D80}, 017502 (2009), [Erratum: Phys.
  Rev.D85,019902(2012)]

\bibitem{Branz:2009}
T.~Branz, T.~Gutsche, V.E. Lyubovitskij, Phys. Rev. \textbf{D80}, 054019 (2009)

\bibitem{Albuquerque:2009}
R.M. Albuquerque, M.E. Bracco, M.~Nielsen, Phys. Lett. \textbf{B678}, 186
  (2009)

\bibitem{Ding:2009}
G.J. Ding, Eur. Phys. J. \textbf{C64}, 297 (2009)

\bibitem{Stancu:2009}
F.~Stancu, J. Phys. \textbf{G37}, 075017 (2010)

\bibitem{Wang:2015a}
Z.g. Wang, Y.f. Tian, Int. J. Mod. Phys. \textbf{A30}, 1550004 (2015)

\bibitem{Anisovich:2015a}
V.V. Anisovich, M.A. Matveev, A.V. Sarantsev, A.N. Semenova, Int. J. Mod. Phys.
  \textbf{A30}(32), 1550186 (2015)

\bibitem{Wang:2009}
Z.G. Wang, Eur. Phys. J. \textbf{C63}, 115 (2009)

\bibitem{Mahajan:2009}
N.~Mahajan, Phys. Lett. \textbf{B679}, 228 (2009)

\bibitem{Aaij:2012pz}
R.~Aaij et~al. (LHCb), Phys. Rev. \textbf{D85}, 091103 (2012)

\bibitem{Lees2014a}
J.P. Lees et~al. (BaBar), Phys. Rev. \textbf{D89}(11), 112004 (2014)

\bibitem{Aaltonen:2011}
T.~Aaltonen et~al. (CDF), Mod. Phys. Lett. \textbf{A32}(26), 1750139 (2017)

\bibitem{Aaij:2016nsc}
R.~Aaij et~al. (LHCb), Phys. Rev. \textbf{D95}(1), 012002 (2017)

\bibitem{Lu:2016a}
Q.F. Lü, Y.B. Dong, Phys. Rev. \textbf{D94}(7), 074007 (2016)

\bibitem{Bhavsar:2018}
T.~Bhavsar, M.~Shah, P.C. Vinodkumar, Eur. Phys. J. \textbf{C78}(3), 227 (2018)

\bibitem{Gonzalez:2015}
P.~González, Phys. Rev. \textbf{D92}, 014017 (2015)

\bibitem{Guo:2014}
P.~Guo, T.~Yépez-Martínez, A.P. Szczepaniak, Phys. Rev. \textbf{D89}(11),
  116005 (2014)

\bibitem{Ke:2013}
H.W. Ke, X.Q. Li, Y.L. Shi, Phys. Rev. \textbf{D87}(5), 054022 (2013)

\bibitem{Ebert2002}
D.~Ebert, R.~Faustov, V.~Galkin, Mod.Phys.Lett. \textbf{A17}, 803 (2002)

\bibitem{Brambilla2006}
N.~Brambilla, eConf \textbf{C0610161}, 004 (2006), \texttt{hep-ph/0702105}

\bibitem{DeFazio:2008}
F.~De~Fazio, Phys. Rev. \textbf{D79}, 054015 (2009), [Erratum: Phys.
  Rev.D83,099901(2011)]

\bibitem{Donald:2012ga}
G.C. Donald, C.~Davies et~al., Phys. Rev. \textbf{D86}, 094501 (2012),
  \texttt{1208.2855}

\bibitem{Liu:2012}
L.~Liu, G.~Moir, Peardon et~al. (Hadron Spectrum), JHEP \textbf{07}, 126 (2012)

\bibitem{Zhu:1998}
S.L. Zhu, Y.B. Dai, Phys. Rev. \textbf{D59}, 114015 (1999)

\bibitem{Beilin:1984}
V.A. Beilin, A.V. Radyushkin, Nucl. Phys. \textbf{B260}, 61 (1985)

\bibitem{Godfrey1985}
S.~Godfrey, N.~Isgur, Phys. Rev. D \textbf{32}, 189 (1985)

\bibitem{Barnes:2005}
T.~Barnes, S.~Godfrey, E.S. Swanson, Phys. Rev. \textbf{D72}, 054026 (2005)

\bibitem{Li:2012vc}
B.Q. Li, C.~Meng, K.T. Chao (2012), \texttt{1201.4155}

\bibitem{Li:2009}
B.Q. Li, K.T. Chao, Phys. Rev. \textbf{D79}, 094004 (2009)

\bibitem{Cao:2012}
L.~Cao, Y.C. Yang, H.~Chen, Few Body Syst. \textbf{53}, 327 (2012)

\bibitem{Segovia:2008}
J.~Segovia, A.M. Yasser, D.R. Entem, F.~Fernandez, Phys. Rev. \textbf{D78},
  114033 (2008)

\bibitem{Eichten1978}
E.~Eichten et~al., Phys. Rev. D \textbf{17}(11), 3090 (1978)

\bibitem{Deng:2016}
W.J. Deng, H.~Liu, L.C. Gui, X.H. Zhong, Phys. Rev. \textbf{D95}(3), 034026
  (2017)

\bibitem{Godfrey:2015}
S.~Godfrey, K.~Moats, Phys. Rev. \textbf{D92}(5), 054034 (2015)

\bibitem{Koma2006}
Y.~Koma, M.~Koma, H.~Wittig, Phys. Rev. Lett \textbf{97}, 122003 (2006)

\bibitem{Kher:2017}
V.~Kher, N.~Devlani, A.K. Rai (2017), \texttt{1704.00439}

\bibitem{Kher:2017b}
V.~Kher, N.~Devlani, A.K. Rai, Chinese Physics C \textbf{Vol. 41,}(No. 9),
  093101 (2017)

\bibitem{Brambilla:2004jw}
N.~Brambilla, A.~Pineda, J.~Soto, A.~Vairo, Rev. Mod. Phys. \textbf{77}, 1423
  (2005)

\bibitem{Brambilla:2014}
N.~Brambilla et~al., Eur. Phys. J. \textbf{C74}(10), 2981 (2014)

\bibitem{Gupta1995}
S.N. Gupta, J.M. Johnson, Phys. Rev. D \textbf{51}(1), 168 (1995)

\bibitem{Hwang1997}
D.S. Hwang, C.~Kim, W.~Namgung, Phys.Lett. \textbf{B406}, 117 (1997),
  \texttt{hep-ph/9608392}

\bibitem{Rai2008}
A.K. Rai, B.~Patel, P.C. Vinodkumar, Phys. Rev. C \textbf{78}(5), 055202 (2008)

\bibitem{Rai2002}
A.K. Rai, R.H. Parmar, P.C. Vinodkumar, J. Phys. G: Nucl. Part. Phys.
  \textbf{28}(8), 2275 (2002)

\bibitem{Eichten:2008}
E.~Eichten, S.~Godfrey, H.~Mahlke, J.L. Rosner, Rev. Mod. Phys. \textbf{80},
  1161 (2008)

\bibitem{Voloshin:2007}
M.B. Voloshin, Prog. Part. Nucl. Phys. \textbf{61}, 455 (2008)

\bibitem{Lakhina2006}
O.~Lakhina, E.S. Swanson, Phys. Rev. D \textbf{74}, 014012 (2006)

\bibitem{VanRoyen1967}
R.~Van~Royen, V.~Weisskopf, Nuovo Cim. \textbf{A50}, 617 (1967)

\bibitem{Braaten1995}
E.~Braaten, S.~Fleming, Phys. Rev. D \textbf{52}(1), 181 (1995)

\bibitem{Hwang1997a}
D.S. Hwang, G.H. Kim, Phys. Rev. D \textbf{55}(11), 6944 (1997)

\bibitem{Ding:2007}
G.J. Ding, J.J. Zhu, M.L. Yan, Phys. Rev. \textbf{D77}, 014033 (2008)

\bibitem{Lu2016}
Q.F. Lu, T.T. Pan, Y.Y. Wang, E.~Wang, D.M. Li, Phys. Rev. \textbf{D94}(7),
  074012 (2016)

\bibitem{Guo:2010a}
F.K. Guo, C.~Hanhart, G.~Li, U.G. Meissner, Q.~Zhao, Phys. Rev. \textbf{D82},
  034025 (2010)

\bibitem{Radford2009}
S.F. Radford, W.W. Repko, M.J. Saelim, Phys. Rev. D \textbf{80}(3), 034012
  (2009)

\bibitem{Segovia:2016}
J.~Segovia, P.G. Ortega, D.R. Entem, F.~Fernández, Phys. Rev. \textbf{D93}(7),
  074027 (2016)

\bibitem{Kwong:1987}
W.~Kwong, P.B. Mackenzie, R.~Rosenfeld, J.L. Rosner, Phys. Rev. \textbf{D37},
  3210 (1988)

\bibitem{Kwong:1988}
W.~Kwong, J.L. Rosner, Phys. Rev. \textbf{D38}, 279 (1988)

\bibitem{Bradley:1980}
A.~Bradley, A.~Khare, Z. Phys. \textbf{C8}, 131 (1981)

\bibitem{Belanger:1987}
G.~Belanger, P.~Moxhay, Phys. Lett. \textbf{B199}, 575 (1987)

\bibitem{Bhaghyesh:2011}
Bhaghyesh, K.B. Vijaya~Kumar, A.P. Monteiro, J. Phys. \textbf{G38}, 085001
  (2011)

\bibitem{Negash:2015}
H.~Negash, S.~Bhatnagar, Int. J. Mod. Phys. \textbf{E25}(08), 1650059 (2016)

\bibitem{Yang:2015cc}
J.H. Yang, S.K. Lee, E.J. Kim, J.B. Choi (2015), \texttt{1506.04481}

\bibitem{Ebert2011}
D.~Ebert, R.~Faustov, V.~Galkin, Eur.Phys.J. \textbf{C71}, 1825 (2011)

\bibitem{Radford:2007}
S.F. Radford, W.W. Repko, Phys. Rev. \textbf{D75}, 074031 (2007)

\bibitem{Ebert:2002}
D.~Ebert, R.N. Faustov, V.O. Galkin, Phys. Rev. \textbf{D67}, 014027 (2003)

\bibitem{Sultan:2014}
M.A. Sultan, N.~Akbar, B.~Masud, F.~Akram, Phys. Rev. \textbf{D90}(5), 054001
  (2014)

\bibitem{Godfrey:1985}
S.~Godfrey, N.~Isgur, Phys. Rev. \textbf{D32}, 189 (1985)

\bibitem{Ebert2010}
D.~Ebert, R.~Faustov, V.~Galkin, Eur.Phys.J. \textbf{C66}, 197 (2010)

\bibitem{Parmar2010}
A.~Parmar, B.~Patel, P.~Vinodkumar, Nucl.Phys. \textbf{A848}, 299 (2010)

\bibitem{Brambilla:2004}
N.~Brambilla et~al. (Quarkonium Working Group) (2004), \texttt{hep-ph/0412158}

\bibitem{Eichten:2002}
E.J. Eichten, K.~Lane, C.~Quigg, Phys. Rev. Lett. \textbf{89}, 162002 (2002),
  \texttt{hep-ph/0206018}

\bibitem{DSouza:2017}
P.P. D'Souza, M.~Bhat, A.P. Monteiro, K.B. Vijaya~Kumar (2017)

\bibitem{Bhaghyesh:2012}
Bhaghyesh, K.B. Vijaya~Kumar, Y.L. Ma, Int. J. Mod. Phys. \textbf{A27}, 1250011
  (2012)

\bibitem{Giannuzzi:2008}
F.~Giannuzzi, Phys. Rev. \textbf{D78}, 117501 (2008)

\bibitem{Laverty:2009}
J.T. Laverty, S.F. Radford, W.W. Repko (2009), \texttt{0901.3917}

\bibitem{Munz:1996}
C.R. Munz, Nucl. Phys. \textbf{A609}, 364 (1996)

\bibitem{Kim:2004}
C.S. Kim, T.~Lee, G.L. Wang, Phys. Lett. \textbf{B606}, 323 (2005)

\bibitem{Ebert:2003b}
D.~Ebert, R.N. Faustov, V.O. Galkin, Mod. Phys. Lett. \textbf{A18}, 601 (2003)

\bibitem{Tornqvist:2004}
N.A. Tornqvist, Phys. Lett. \textbf{B590}, 209 (2004)

\bibitem{Braaten:2007}
E.~Braaten, M.~Lu, Phys. Rev. \textbf{D76}, 094028 (2007)

\bibitem{Albaladejo:2017}
M.~Albaladejo, F.K. Guo, C.~Hanhart, U.G. Meißner, J.~Nieves, A.~Nogga,
  Z.~Yang, Chin. Phys. \textbf{C41}(12), 121001 (2017)

\bibitem{Hanhart:2007}
C.~Hanhart, {\relax Yu}.S. Kalashnikova, A.E. Kudryavtsev, A.V. Nefediev, Phys.
  Rev. \textbf{D76}, 034007 (2007)

\bibitem{Zhao:2013}
C.W. Zhao, G.~Li, X.H. Liu, F.L. Shao, Eur. Phys. J. \textbf{C73}, 2482 (2013)

\bibitem{Liu:2009fe}
X.~Liu, Z.G. Luo, Z.F. Sun, Phys. Rev. Lett. \textbf{104}, 122001 (2010)

\bibitem{Guo:2010}
F.K. Guo, C.~Hanhart, G.~Li, U.G. Meissner, Q.~Zhao, Phys. Rev. \textbf{D83},
  034013 (2011)

\bibitem{Li:2007}
G.~Li, Q.~Zhao, Phys. Lett. \textbf{B670}, 55 (2008)

\bibitem{Li:2011a}
G.~Li, Q.~Zhao, Phys. Rev. \textbf{D84}, 074005 (2011)

\bibitem{Afonin:2016}
S.S. Afonin, I.V. Pusenkov, EPJ Web Conf. \textbf{125}, 04006 (2016)

\end{thebibliography}

\end{multicols}

\vspace{-1mm}
\centerline{\rule{80mm}{0.1pt}}
\vspace{2mm}

\begin{multicols}{2}

\end{multicols}

\clearpage

\end{document}